\documentclass{jpp}
\usepackage{graphicx}

\usepackage[utf8]{inputenc}
\usepackage[T1]{fontenc}
\usepackage{amsmath}
\usepackage{amssymb}
\usepackage{xcolor}
\usepackage{bigints}
\usepackage{dsfont}
\usepackage{subcaption}
\usepackage[colorlinks=true]{hyperref}

\shorttitle{Omnigenous umbilic stellarators}
\shortauthor{Gaur et al.}

\newcommand{\NFP}{n_{\rm{FP}}}

\newcommand{\bm}[1]{\boldsymbol{#1}}

\title{Omnigenous umbilic stellarators}

\author{R. Gaur\aff{1}\corresp{\email{rgaur@terpmail.umd.edu}}, D. Panici\aff{1}, T. M. Elder\aff{2, 3}, M. Landreman\aff{3},\\
 K. E. Unalmis\aff{1}, Y. Elmacioglu\aff{1}, D. Dudt\aff{5}, R. Conlin\aff{3}, E. Kolemen\aff{1, 6}}
\date{March 2025}
  
\affiliation{
\aff{1} Princeton University, Princeton, 08540, NJ, USA
\aff{2} Max Planck Institute for Plasma Physics, 17491, Greifswald, Germany,
\aff{3} Institute for Research in Electronics and Applied Physics, University of Maryland, College Park, 20740, MD, USA
\aff{5} Thea Energy, USA
\aff{6} Princeton Plasma Physics Laboratory, Princeton, 08540, NJ, USA
}

\begin{document}
\maketitle

\begin{abstract}
To better understand the dependence of the magnetic field structure in the plasma edge on the plasma boundary shape, in the context of X-point and island divertor designs, we define and develop a class of stellarators called umbilic stellarators. These equilibria are characterized by a single continuous high-curvature edge on the plasma boundary that goes around multiple times toroidally before meeting itself. We develop a technique that allows us to simultaneously optimize the plasma boundary along with a curve lying on the boundary on which we impose a high curvature while imposing omnigenity --- a property of the magnetic field that ensures trapped particle confinement throughout the plasma volume. We find that umbilic stellarators naturally tend to favor piecewise omnigenity instead of omnigenity with a specific helicity. After generating omnigenous umbilic stellarators, we design coil sets for some of them and explore the field line structure in the edge and its sensitivity to small fluctuations in the plasma. Finally, using single-stage optimization, we simultaneously modify the plasma and coil shape and propose an experiment to modify an existing tokamak to a finite-$\beta$ stellarator using this technique and explore a potentially simpler way to convert a limited tokamak into a diverted stellarator.
\end{abstract}

\section{Introduction}
Stellarators~\citep{spitzer1958stellarator} are toroidal nuclear fusion devices that are used to study magnetic confinement fusion. Unlike the axisymmetric tokamaks, stellarators possess discrete toroidal symmetry. The lack of continuous toroidal symmetry offers considerably greater design freedom, at the expense of trapped particle confinement, a quality guaranteed in axisymmetry through Noether's theorem~\citep{Noether1918}. Therefore, we seek stellarators with trapped particle confinement, a property known as omnigenity. 

Omnigenous magnetic fields ensure zero bounce-averaged radial drift \citep{hall1975three,cary1997omnigenity}, of which both axisymmetry and quasisymmetry are subsets \citep{boozer_transport_1983,nuhrenberg_quasi-helically_1988,dudt2024magnetic}. Recent advances in stellarator optimization have generated omnigenous configurations with excellent neoclassical confinement levels\citep{landreman_magnetic_2022,goodman2023constructing}. However, these configurations are typically only calculated up to the Last Closed Flux Surface (LCFS). To develop a stellarator Fusion Power Plant (FPP), it is essential to also predict and optimize the magnetic field structure outside the LCFS~\citep{parra2024flexible}. This is necessary to separate the high-temperature plasma from the Plasma Facing Components (PFCs) while efficiently disposing of fusion by-products like helium ash and impurities such as heavier ions. 

The structure outside the plasma boundary depends on the boundary shape, so changing the boundary shape is a potential way to obtain a desirable fieldline structure in the edge. Indeed, divertors~\citep{burnett1958divertor} have been greatly successful in tokamaks and offer significantly improved performance over limiters~\citep{wagner1982regime}. Depending on the boundary shape and coil positions, one can design various divertor configurations in tokamaks such as the single-null, double-null, snowflake~\citep{ryutov2007geometrical}, and super-X~\citep{valanju2009super}.These configurations utilize poloidal and divertor coils to create and shift an X point --- a region in tokamaks where the poloidal field disappears.

Stellarators have utilized three distinct divertor configurations~\citep{baderprogress}: helical, resonant, and non-resonant divertors. Helical divertors were used in heliotron/torsatron concepts such as the Heliac device~\citep{harris1985flexible} and LHD~\citep{ohyabu1992helical}. These divertors work by creating a stochastic region with edge surface layers and X-points, which the magnetic field lines cross before reaching the scrape-off layer(SOL). The coil design and magnetic configuration in the edge surface layers is complex and difficult to predict~\citep{ohyabu1994large}. Moreover, applicability of this concept to omnigenous stellarators with non-helical symmetry still remains uncertain~\citep{baderprogress}, posing challenges for their design and optimization.

Resonant divertors, used in W7-AS~\citep{grigull2001first,feng2002transport} and W7-X~\citep{pedersen2019first}, work by leveraging island formation and placing divertors between the X and O points of the islands. The island structure depends on the pitch of the field line in the boundary, whereas the connection length depends directly on the number of X points and inversely on the magnetic field line pitch and the magnetic shear~\citep{feng2011comparison} making them useful for low magnetic shear stellarators. 

Non-resonant divertors use a more complicated design to create a stochastic field in the edge, but similar to resonant divertors, they can be used to spread the heat load and improve plasma exhaust rejection. These divertors are also more resilient to changes in the magnetic field structure~\citep{bader2017hsx}, but the magnetic field structure in the chaotic field region is complicated to calculate and requires further understanding. Recent work~\citep{boozer2018simulation, punjabi2022magnetic, garcia2023exploration, davies2025topology} has further explored the possibility of using non-resonant divertors in stellarators.

Motivated by the need to understand how boundary shaping affects fieldline structure in the edge and X-point and island divertor design, in this paper, we study the umbilic stellarator (US) as a method of achieving control over the edge magnetic field topology, and generate umbilic stellarators for vacuum and finite-$\beta$ cases\footnote{where $\beta = 2\mu_0 p/B^2$ is the ratio of the plasma to the magnetic pressure}. Umbilic stellarators are characterized by regions of high curvature on the boundary and how these high-curvature regions are positioned. We develop a technique that allows us to obtain umbilic stellarators by simultaneously optimizing a curve and the plasma boundary, and imposing a high curvature along this curve which creates a ridge along the plasma boundary. Additionally, we ensure that all equilibria have reasonable omnigenity, ensuring low neoclassical transport. Finally, we propose an experiment to modify an existing tokamak to a stellarator using our optimization process and build it using umbilic coils, which are helically wound coils, similar to divertor coils, with fractional helicity that carry a small current (compared to the plasma current), and follow an umbilic edge on the plasma boundary. We show how an umbilic coil can create a high-curvature umbilic region in finite-pressure equilibria, acting as an alternative to modular coils, to create a ridge on the plasma boundary.

In~\S\ref{sec:2}, we explain the ideal MHD equilibrium. In~\S\ref{sec:3}, we introduce the umbilic boundary shape, how to parametrize them, and how we approximate them with~\texttt{DESC}, and how to solve the ideal MHD equations throughout the volume enclosed by the approximated shape. In~\S\ref{sec:eq-only-optimization}, we present a set of optimized vacuum($\beta = 0$) and finite-$\beta$ stellarators and analyze their properties such as omnigenity.  To understand the magnetic field structure in the edge, we design modular coils for the vacuum equilibrium and trace the fieldlines throughout the volume. In order to assess the robustness of the edge magnetic field configuration, we introduce a dummy current source at the magnetic axis to emulate a plasma fluctuation. In~\S\ref{sec:HBT-coil}, we take the Columbia HBT-EP plasma shape, modify it into an umbilic stellarator without changing the coilset used for the original tokamak, just by adding an umbilic coil. We study cases where the umbilic coil carries current in the same (co-$I$) and opposite (counter-$I$) direction as the plasma current $I$, study the effect of the umbilic current on edge iota, and solve a single stage problem for the co-$I$ case, by simultaneously modifying the plasma boundary and umbilic coil shape. In~\S\ref{sec:fin}, we conclude our study by highlighting existing limitations and exploring potential avenues for overcoming them.

\section{Ideal MHD equilibrium}
\label{sec:2}
In this section, we briefly explain how we define and calculate an ideal MHD equilibrium in a stellarator. We then explain how to locally vary the gradients of that equilibrium. 

A divergence-free magnetic field $\bm{B}$ can be written in the Clebsch form~\citep{d2012flux}
\begin{equation}
    \bm{B} =   \nabla \psi \times \nabla\alpha,
    \label{eqn:Div-free-B2}
\end{equation}
We will focus on solutions whose magnetic field lines lie on closed nested toroidal surfaces, known as flux surfaces. We label these surfaces using the enclosed toroidal flux $\psi$. On each flux surface, the lines of constant $\alpha$ coincide with the magnetic field lines. Thus, $\alpha$ is known as the field line label. We define $\alpha = \theta_{\rm{PEST}} - \iota \zeta$, where $\theta_{\rm{PEST}}$ is the $\mathrm{PEST}$ straight field line angle, $\zeta$ is the cylindrical toroidal angle, and 
\begin{equation}
    \iota = \frac{\bm{B}\cdot \bm{\nabla}\theta_{\rm{PEST}}}{\bm{B}\cdot \bm{\nabla}\zeta},
\end{equation}
is the pitch of the magnetic field lines on a flux surface, known as the rotational transform. Using the Clebsch form of the magnetic field, we solve the ideal MHD force balance equation
\begin{equation}
    \bm{j} \times \bm{B} = \bm{\nabla} p,
    \label{eqn:ideal-MHD-force-balance}
\end{equation}
where the plasma current $\bm{j} = (\bm{\nabla} \times \bm{B})/\mu_0$ from Ampere's law, $p$ is the plasma pressure and $\mu_0$ is the vacuum magnetic permeability. Unlike an axisymmetric case, for stellarators we have to solve~\eqref{eqn:ideal-MHD-force-balance} as an optimization problem. We achieve this with the $\texttt{DESC}$~\citep{dudt2020desc, panici2023desc, conlin2023desc, dudt2023desc} stellarator optimization suite through minimization of $F \equiv  \bm{j} \times \bm{B} - \bm{\nabla} p$. $\texttt{DESC}$ can simultaneously solve an equilibrium while optimizing for multiple objectives such as MHD stability, quasisymmetry and more, without re-solving the equilibrium force balance equation~\eqref{eqn:ideal-MHD-force-balance} for each optimization variable. To reduce the computational cost further, we also use the fact that stellarators possess a discrete toroidal symmetry such that every stellarator comprises several identical sections. The number of identical sections is denoted by the field period $\NFP$. Therefore, the force balance problem is only solved in a single field period.
In the following sections, we utilize~\texttt{DESC} to optimize for omnigenous umbilic stellarators.

\section{Umbilic stellarators}
\label{sec:3}
\label{sec:UToL}
Umbilic stellarators draw inspiration from an umbilic bracelet (or torus)~\citep{zeeman_umbilic_1976}, a 3D shape with a cross-section that is a deltoid curve, a three-cusped hypocycloid which rotates by $\Delta \theta = 2\pi/3$ poloidally after a single toroidal turn $\Delta \zeta = 2\pi$. The sharp edge lies exactly on the cusp. Therefore, the sharp edge has to go around three times toroidally before meeting itself. A stellarator generated using an umbilic boundary is potentially advantageous for divertor designs, offering a continuous divertor option with a long connection length. For a smooth magnetic field, the sharp edge of such a boundary would be impossible for a magnetic field line. Therefore, the fieldline pitch on the boundary matches the umbilic topology. That is, $\iota = \NFP m/n = 1/3$, where $m, n \in \mathbb{Z}$ corresponding to the poloidal and toroidal motion of the fieldline and $\NFP$, the field period, is the measure of discrete symmetry of the stellarator. For a general umbilic shape, $m/n$ can attain rational values besides $1/3$. Further details are provided in this section.

\subsection{Defining an umbilic surface and edge}
\label{subsec:UToL-parametrization}
\begin{figure}
    \centering
    \begin{subfigure}[b]{0.48\textwidth}
    \centering
        \includegraphics[height=5cm, trim={2mm 2mm 2mm 4mm}, clip]{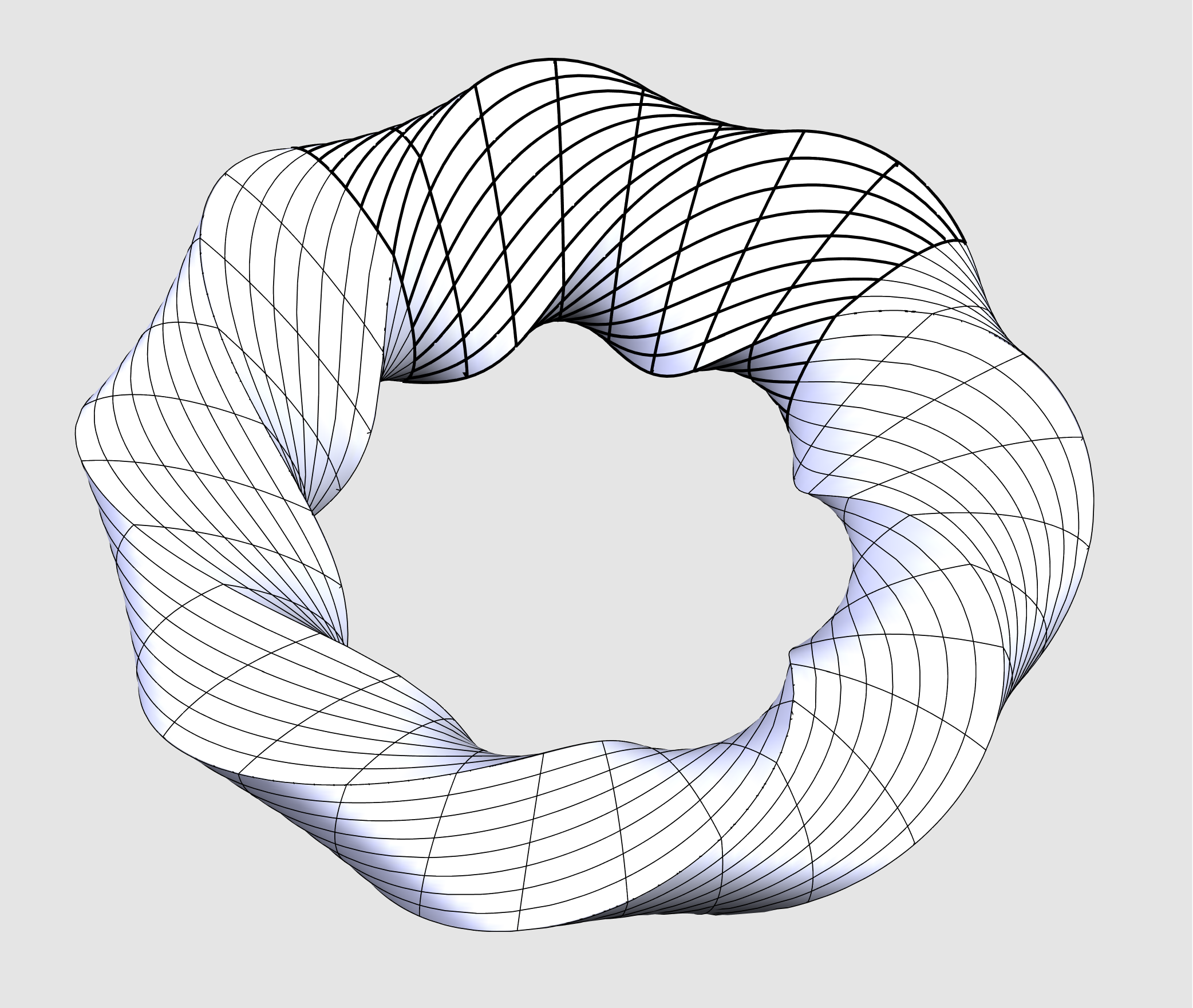}
        \caption{US432 full surface}
    \end{subfigure}
    \begin{subfigure}[b]{0.48\textwidth}
        \centering
        \includegraphics[height=5cm, trim={2mm 50mm 2mm 50mm}, clip]{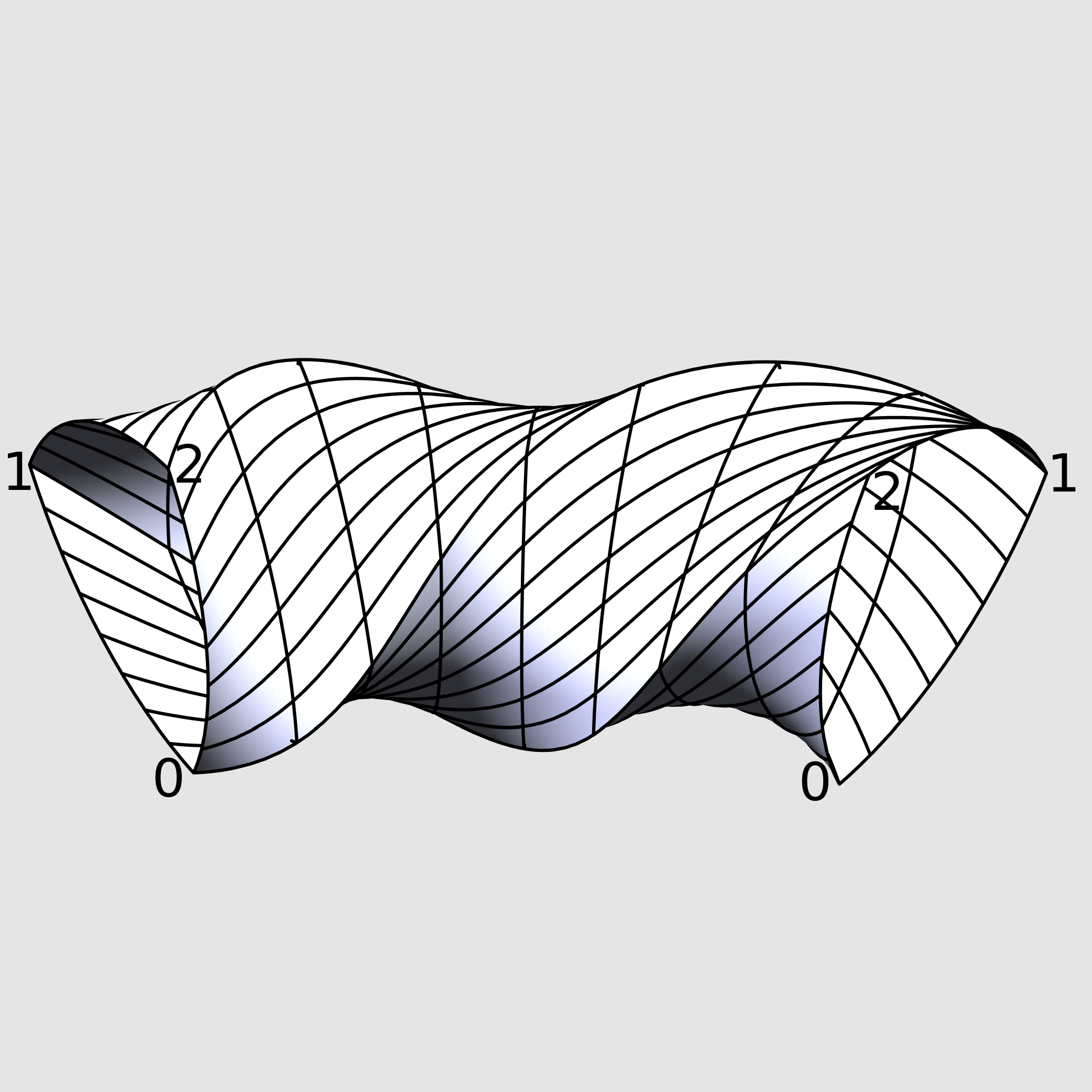}
        \caption{US432 single field period}
    \end{subfigure}
    \caption{An umbilic boundary design with $\NFP=4, n=3, m=2, \varrho = 0.344$. The rest of the parameters $r_j=0$. In \textit{(a)} we show the full shape and in \textit{(b)} we present a single field period from the parametrization. A point $p_i \in \{0, 1, 2\}$ corresponding to the sharp edge at each toroidal cross section connects to$\mod(p_i+m, 3)$ after a single field period. The sharp edge defined by $R_{\mathrm{US}}(\theta= 0, \zeta), Z_{\mathrm{US}}(\theta=0, \zeta)$ meets itself after $n = 3$ toroidal turns.}
    \label{fig:UToLS_parametrization}
\end{figure}

The initial shape of a stellarator symmetric umbilic-torus-like surface $\partial V$ is parameterized as a rotating polygon-like with convex sides
\begin{equation}
\begin{split}
R_{\mathrm{US}} & =  R_0 + 2\varrho \cos\left(\frac{\pi}{2n}\right) \cos\left(\frac{\theta + \frac{\pi}{n} (2 \lfloor \frac{n \theta}{2\pi} \rfloor + 1)}{2} 
+ \frac{m \NFP \zeta + r_1 \sin(\NFP \zeta) + r_4 \sin(2 \NFP \zeta)}{n} \right) \\
& - \varrho \cos\left(\frac{\pi}{n}(2 \Big \lfloor \frac{n \theta}{2\pi} \Big \rfloor + 1) + \frac{m 
\NFP \zeta + r_1 \sin(\NFP \zeta) + r_4 \sin(2 \NFP \zeta)}{n}\right)  + r_2 \cos(\NFP \zeta)
\end{split}
\label{eqn:umbilic-surface-R}
\end{equation}

\begin{equation}
\begin{split}
Z_{\mathrm{US}} &= 2\varrho \cos\left(\frac{\pi}{2n}\right) \sin\left(\frac{\theta + \frac{\pi}{n} (2 \lfloor \frac{n \theta}{2\pi} \rfloor + 1)}{2} 
+ \frac{m \NFP \zeta + r_5 \sin(\NFP \zeta)}{n} \right) \\
& - \varrho \sin\left(\frac{\pi}{n}(2 \Big\lfloor \frac{n \theta}{2\pi} \Big\rfloor + 1) 
+ \frac{m \NFP \zeta + r_5 \sin(\NFP \zeta)}{n}\right) + r_3 \sin( \NFP \zeta)
\end{split}
\label{eqn:umbilic-surface-Z}
\end{equation}
where the major radius $ R_0 = 1$ throughout this paper, unless stated otherwise, $\varrho$ is the minor radius of the umbilic-torus-like surface, $\theta$ is the poloidal angle used in~\texttt{DESC}, $\zeta$ is the cylindrical toroidal angle, and $\lfloor ... \rfloor$ is the floor function, defined such that
\[
\lfloor x \rfloor =
\begin{cases}
x, & \text{if } x \in \mathbb{Z} \\
N-1, & \text{if } N-1 < x < N \text{ and } x \notin \mathbb{Z}, \ N \in \mathbb{Z} 
\end{cases}
\]
In the surface parametrization, we also define $n$ as the umbilic factor, which defines the number of sides in the analytical parametrization, $m$ is an integer that determines the poloidal motion of the sharp umbilic edge, and $\NFP$ is the field period. The periodicity of the edge is $\NFP (m/n)$ --- the edge closes on itself after $n/\gcd(n, \NFP)$ toroidal turns, where $\gcd$ is the greatest common divisor. The parameters $r_1, r_4, r_5$ define the convexity/concavity of a cross-section and how it changes toroidally, $r_4$ determines the magnetic axis torsion and $r_2$ is another parameter that determines the shape of the axis. The shape of such a boundary, the naming convention we shall use, and periodicity of the edge are illustrated in figure~\ref{fig:UToLS_parametrization}.

The floor function creates the sharp umbilic edge, while the rest of the parameters grant us enough freedom that we can start with a boundary shape that has good quasisymmetry and allows us to create an umbilic edge with any low-order rational periodicity $n_{\rm{FP}} (m/n)$. It is also stellarator symmetric, i.e., $R_{\mathrm{US}}(\theta, \zeta) = R_{\mathrm{US}}(-\theta, -\zeta), Z_{\mathrm{US}}(\theta, \zeta) = -Z_{\mathrm{US}}(-\theta, -\zeta)$.

We also define a separate grid on the boundary $\rho = 1$, the umbilic edge, on which a sharp curvature is imposed, defined by values $(\hat{\theta}, \zeta)$ so that
\begin{equation}
    \hat{\theta} = \frac{m \NFP \zeta + \sum_{k}^{N_{\mathrm{u}}} a_{k} \sin(\NFP \zeta)}{n}, \quad \gcd(m, n) = 1,
    \label{eqn:UToL-edge}
\end{equation}
For the surface parametrization, $\theta = 0$ corresponds to the umbilic edge --- initial $\hat{\theta}$ is calculated along this edge but it can change freely during optimization as the optimizer varies the coefficients $a_k$. Note that $\theta$ is the DESC poloidal angle throughout this paper, unless stated otherwise.

However, creating and solving with a perfectly sharp edge is difficult with a fully spectral solver. Therefore, in the next section, we will explain how we approximate umbilic boundary shapes with the~\texttt{DESC} code.

\subsection{Approximating umbilic shapes with \texttt{DESC}}
\label{subsec:fitting-UToLs}
\texttt{DESC} uses a Fourier-Zernike spectral representation to solve the force balance equation in the $(\rho, \theta, \zeta)$ coordinate system, where $\rho = \sqrt{\psi/\psi_b},\, \theta = \theta_{\rm{PEST}} - \Lambda(\rho, \theta, \zeta), \, \zeta = \phi$, the cylindrical toroidal angle.
The problem is defined in a cylindrical coordinate system $(R, \zeta, Z)$ by decomposing it into Fourier-Zernike spectral bases as shown below (See \citep{panici2023desc} for more details)
\begin{equation}
    R(\rho,\theta,\zeta) = \sum_{m=-M,n=-N,l=0}^{M,N,L} R_{lmn} \mathcal{Z}_l^m (\rho,\theta) \hat{F}^n(\zeta)
\end{equation}
\begin{equation}
Z(\rho,\theta,\zeta) = \sum_{m=-M,n=-N,l=0}^{M,N,L} Z_{lmn} \mathcal{Z}_l^m (\rho,\theta) \hat{F}^n(\zeta)
\end{equation}
where $l, m$, and $n$ are the radial, poloidal, and toroidal mode numbers, whereas $L, M$, and $N$ define the largest values of $l, m$, and $n$, respectively, and $\mathcal{Z}, \hat{F}$ are Zernike and Fourier basis functions. These parameters define the resolution of a\texttt{DESC} equilibrium. 

\begin{figure}
    \centering
    \includegraphics[height=6cm, trim={2mm 0mm 0mm 0mm}, clip]{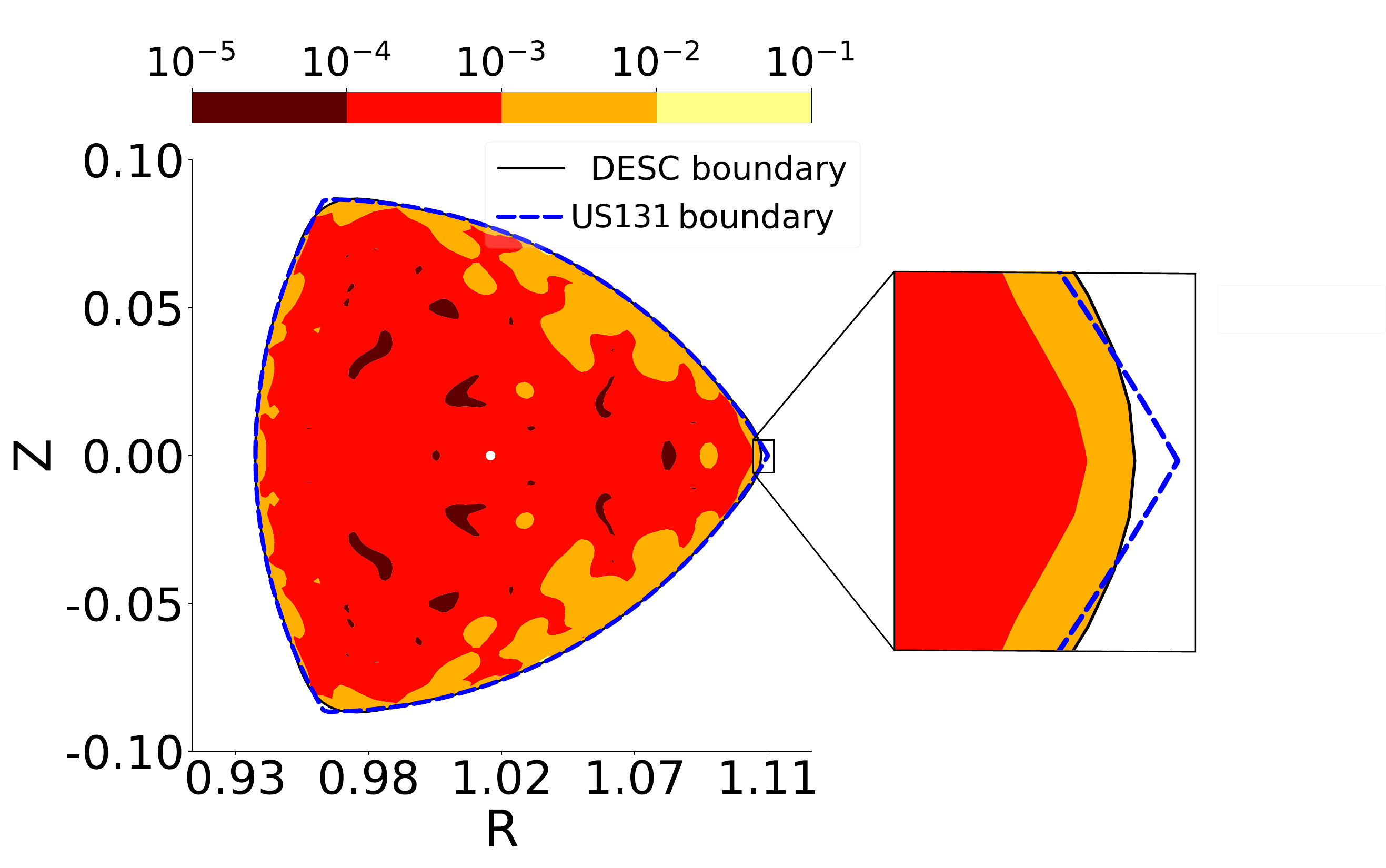}
    \caption{Contour plot of the flux surface average force balance error $|F|$ plotted on the cross section $\zeta = 0$ of a vacuum~\texttt{DESC} equilibrium created using a US131 boundary with $\rho = 0.1, r_1=0.133, r_5=0.1$. The actual boundary (in solid black) is obtained after fitting a Fourier-Zernike series to the true umbilic boundary(in dashed blue).The resolution of the~\texttt{DESC} equilibrium is $L=M=N=14$ and the resolution of the umbilic curve $N_{\mathrm{u}} = 12$.}
    \label{fig:normF_UToL131}
\end{figure}

To solve the ideal MHD equation inside the umbilic shape, we have to first fit the Fourier-Zernike basis at $\rho = 1$ to the umbilic boundary, which requires solving a set of linear equations.

\begin{eqnarray}
\centering
\left[\begin{array}{cc}
    \sum_{m=-M,n=-N,l=0}^{M,N,L} R_{lmn} \mathcal{Z}_l^m (\rho = 1,\theta) \hat{F}^n(\zeta)\\[2mm]
    \sum_{m=-M,n=-N,l=0}^{M,N,L} Z_{lmn} \mathcal{Z}_l^m (\rho = 1,\theta) \hat{F}^n(\zeta)
\end{array}\right] 
= 
\left[\begin{array}{cc}
    R_{\mathrm{US}}(\theta, \zeta)\\[2mm]
    Z_{\mathrm{US}}(\theta, \zeta)
\end{array}\right], 
\end{eqnarray}
to find $R_{lmn}, Z_{lmn}$ so that the boundary shape matches with the umbilic parametrization.
Similarly, to obtain $a_k$, we parameterize the umbilic edge using $R_{\mathrm{curve}} = R(\rho=1, \theta=0, \zeta), Z_{\mathrm{curve}} = Z(\rho = 1, \theta=0, \phi)$, and perform an inverse Fourier transform to obtain $a_k$.
Note that upon fitting a spectral basis, the umbilic edge is smoothed out. After representing the umbilic shape in Fourier-Zernike basis, we solve the ideal MHD equation inside the volume. Figure~\ref{fig:normF_UToL131} illustrates this process of fitting and solving force balance in a US131 boundary shape.

The sharpness of the umbilic edge increases as we increase the resolution of the fit surface and the agreement between $\texttt{DESC}$ and the umbilic parametrization improves. Alternatively, one may use a double Chebyshev~\citep{fortunato2024high} instead of a double Fourier representation to improve the spectral fit, local basis methods such as finite-element to represent and solve the equilibrium, or use a non-linear transformation to map the circular torus in Fourier-Zernike basis to an umbilic shape, similar to the transformation used for circular tokamaks by~\citet{webster2009magnetohydrodynamic}. However, using a new basis and performing equilibrium optimization is beyond the scope of this work. 

In the following section, we discuss how, beginning with the parametrization of the umbilic surface, we can optimize for omnigenous umbilic stellarators.

\subsection{Optimizing an umbilic equilibrium with~\texttt{DESC}}
\label{subsec:optimizing-UToLs}

As we noted in the previous section and figure~\ref{fig:normF_UToL131}, after fitting a Fourier-Zernike basis to the umbilic boundary, it becomes smooth because approximating a sharp boundary with a double-Fourier representation is usually difficult. Therefore, we have to modify the optimization process for an umbilic equilibrium to maintain a high curvature, instead of a perfectly sharp umbilic edge. Additionally, for these shapes to be practically feasible fusion reactors, we also need good omnigenity. Therefore, we use the omnigenity and effective ripple objectives to improve the omnigenity throughout the volume.

After fitting to a Fourier-Zernike basis, the high-curvature edge, in general, does not align with the field line. In the unoptimized shape defined in~\eqref{eqn:umbilic-surface-R} and~\eqref{eqn:umbilic-surface-Z}, the $\iota$ is typically too small because of weak shaping. Therefore, we have to optimize separately for a boundary rotational transform to be close $\NFP m/n$. Also, to ensure that the umbilic edge curve matches with the field line labeled $\alpha=0$, we enforce a strong negative curvature $\kappa_{2, \rho}$ along the umbilic edge defined by~\eqref{eqn:UToL-edge}. The definitions of the principal curvatures are given in Appendix~\ref{sec:appendix-curvature}. The overall objective function is
\begin{eqnarray}
   \mathcal{F}_{\mathrm{stage-one}} =  w_{A}\, f_{\mathrm{aspect}}^2  + w_{\iota}\, f_{\rm{iota}}^2 +  w_{O}\, f_{\mathrm{om}}^2 +  w_{U}\, f_{\mathrm{umbilic}}^2 + w_R\, f_{\mathrm{ripple}}^2,
   \label{eqn:overall-objective}
\end{eqnarray}
where $f_{x}$ are various objectives on the right side for aspect ratio, boundary rotational transform, omnigenity, high curvature along the umbilic edge, and effective ripple and $w_A, w_{\iota}, w_{O}, w_{U}, w_{R}$ are weights used with each objective function. The exact definitions of these objectives are provided in appendix~\ref{sec:objectives-FoMs-defn}. At each iteration of a~\texttt{DESC} optimization, $\mathcal{F}_{\mathrm{stage-one}}$ is minimized while satisfying~\eqref{eqn:ideal-MHD-force-balance}. Mathematically,
\begin{equation}
    \min \mathcal{F}_{\mathrm{stage-one}}(\hat{\bm{p}}), \qquad \textrm{s.t.} \quad \bm{\nabla}\left(\mu_0 p + \frac{B^2}{2}\right) - \bm{B}\cdot\bm{\nabla}\bm{B} =  0,\quad  \psi_{\mathrm{b}} = \psi_{\mathrm{b}0},
\end{equation}
where $\psi_{\mathrm{b}0}$ is the user-provided enclosed toroidal flux by the boundary and $\hat{\bm{p}} \in \{R_{\mathrm{b}, mn}, Z_{\mathrm{b}, mn}, a_k\}$ comprises parameters that determine the boundary shape and the position of the umbilic curve on the boundary. With the objective $\mathcal{F}_{\mathrm{stage-one}}$ defined, we run~\texttt{DESC} on a single NVIDIA A100 GPU. A single optimization takes less than $90$ minutes.

\section{Omnigenous umbilic stellarators}
\label{sec:eq-only-optimization}
In this section, we will develop two omnigenous umbilic equilibria using the process explained in the previous sections. The first equilibrium will be a single field period vacuum stellarator with $n=3, m = 1$ umbilic topology for which we also design coils, and the second equilibrium will be a two field period finite-$\beta$ stellarator with $n=5, m=2$ umbilic topology.
\subsection{US131($\NFP = 1, n=3, m =1$) omnigenous vacuum stellarator}
\label{subsec:vacuum-UToL}
\begin{figure}
    \centering
    \hspace*{-4mm}
    \begin{subfigure}[b]{0.44\textwidth}
    \centering
        \includegraphics[width=\textwidth, trim={2mm 6mm 0mm 6mm}, clip]{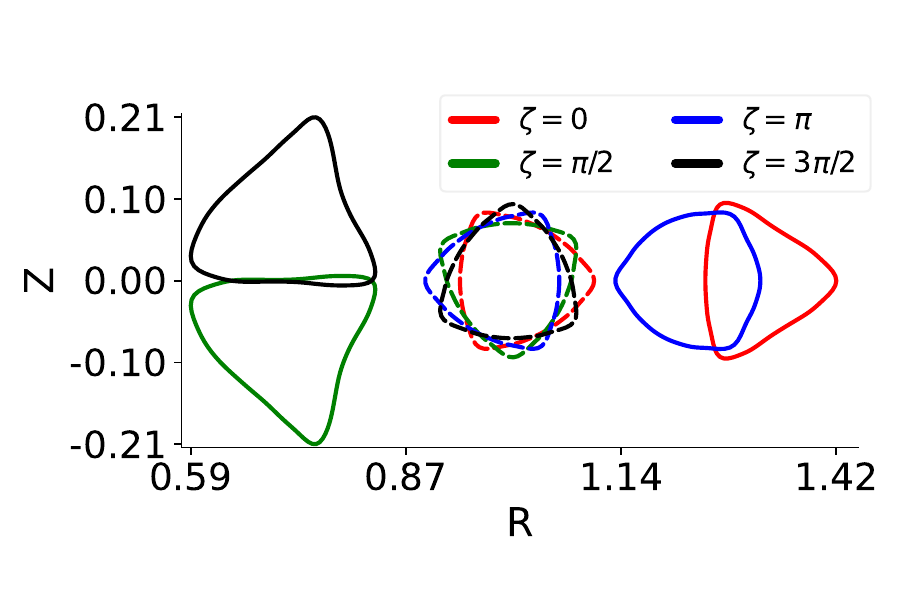}\\[-2mm]
        \caption{Cross section comparison}
    \end{subfigure}
    \hspace*{-5.2mm}
    \begin{subfigure}[b]{0.29\textwidth}
        \centering
        \includegraphics[width=0.96\textwidth, trim={0mm 0mm 3mm 3mm}, clip]{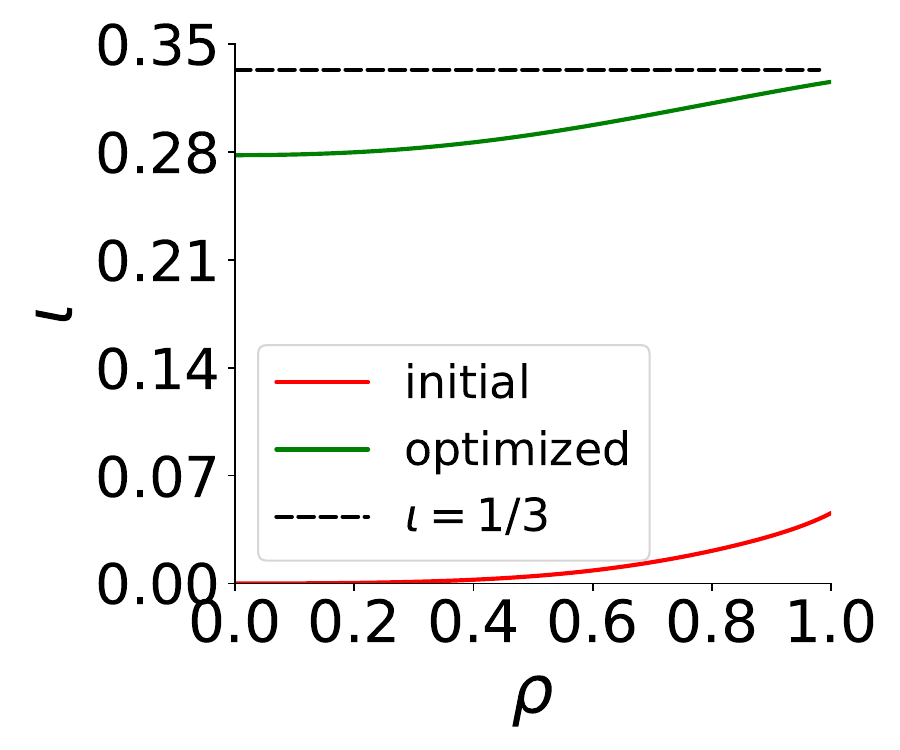}\\[-1mm]
        \caption{Rotational transform}
    \end{subfigure}
    \hspace*{-4.5mm}
    \begin{subfigure}[b]{0.305\textwidth}
        \centering
        \includegraphics[width=0.96\textwidth, trim={0mm 0mm 5mm 3mm}, clip]{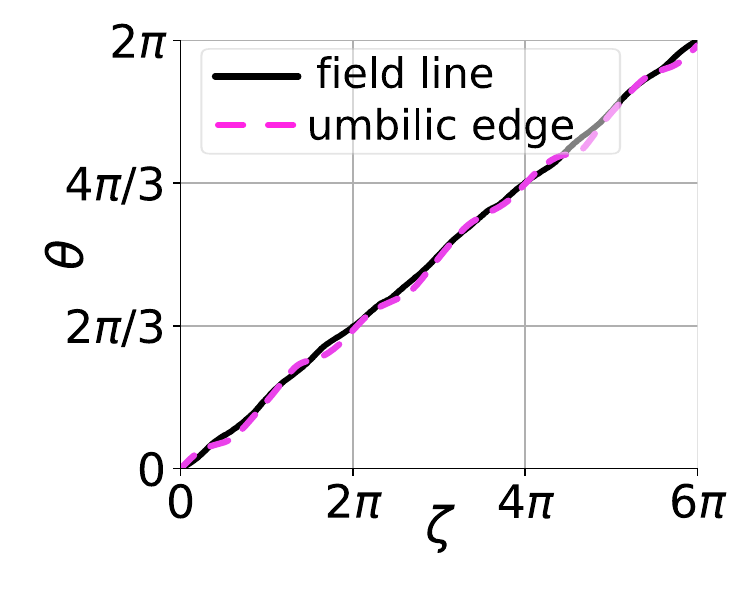}\\[-1.5mm]
        \caption{Alignedness$(\rho = 1.0)$}
    \end{subfigure}
    \caption{US131 optimization results showing a significant increase in magnetic axis torsion in figure~\textit{(a)} from the initial (dotted) to the optimized (solid) boundary. This leads to an increase in the rotational transform, as shown in figure~\textit{(b)} such that $\iota(\rho = 1)$  approaches  $1/3$. The shaping also leads the fieldlines to align with the umbilic edge in the optimized equilibrium as shown in figure~\textit{(c)}.}
\label{fig:vacuum-result}
\end{figure}

\begin{figure}
    \centering
    \begin{subfigure}[b]{0.33\textwidth}
    \centering
        \includegraphics[width=\textwidth, trim={2mm 3mm 2mm 2mm}, clip]{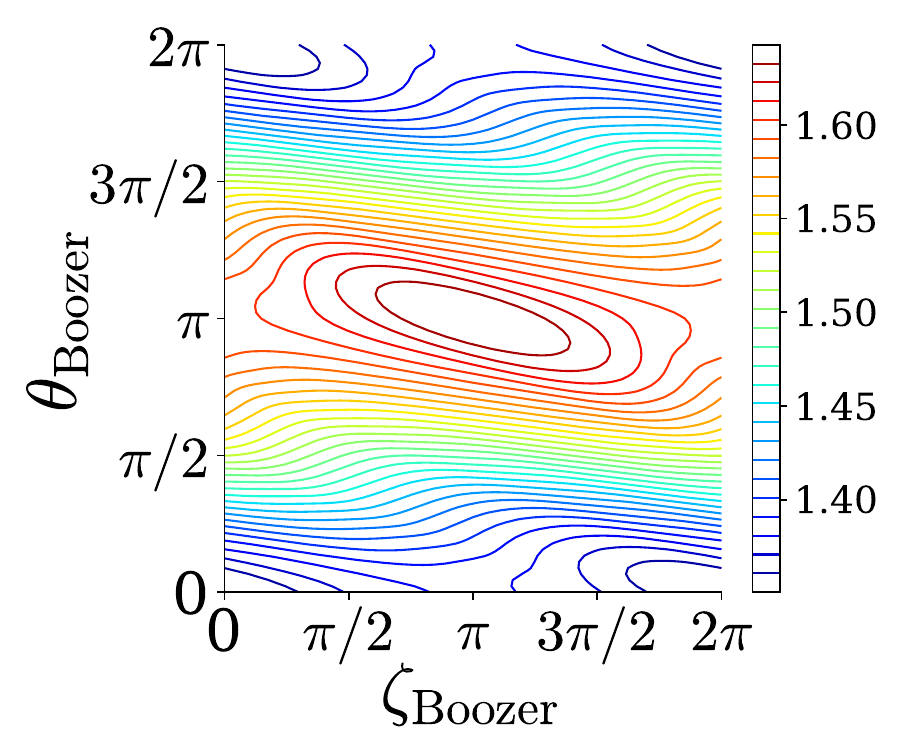}
        \caption{Initial B ($\rho = 1$)}
    \end{subfigure}
    \hspace*{-2mm}
    \begin{subfigure}[b]{0.33\textwidth}
        \centering
        \includegraphics[width=1.0\textwidth, trim={0mm 2mm 0mm 2mm}, clip]{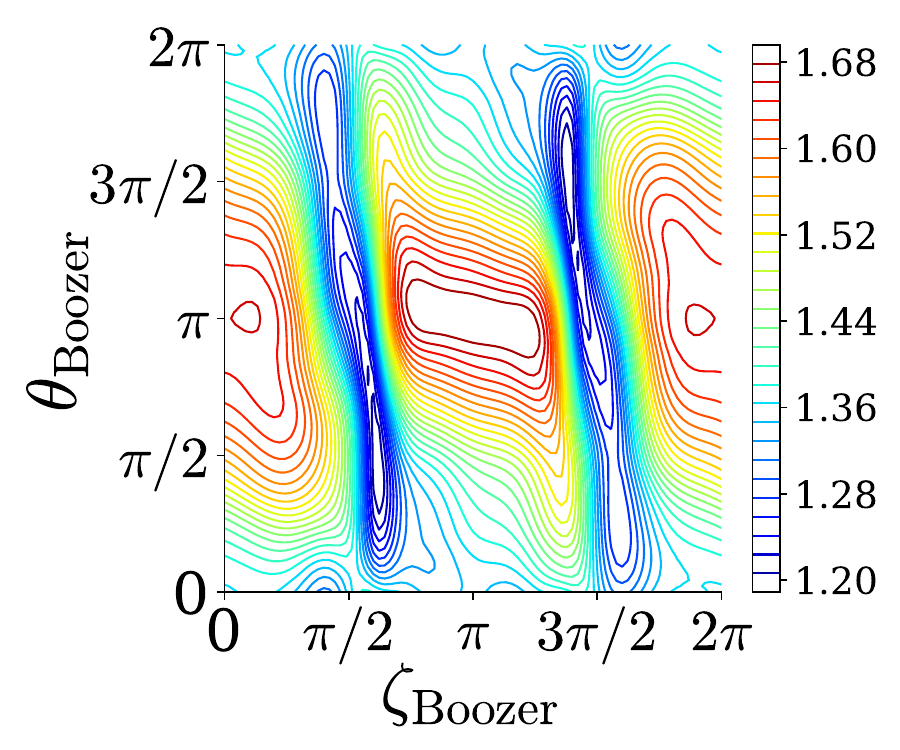}
        \caption{Optimized B ($\rho = 1$)}
    \end{subfigure}
    \hspace*{-2mm}
    \begin{subfigure}[b]{0.33\textwidth}
        \centering
        \includegraphics[width=1.0\textwidth, trim={0mm 0mm 3mm 3mm}, clip]{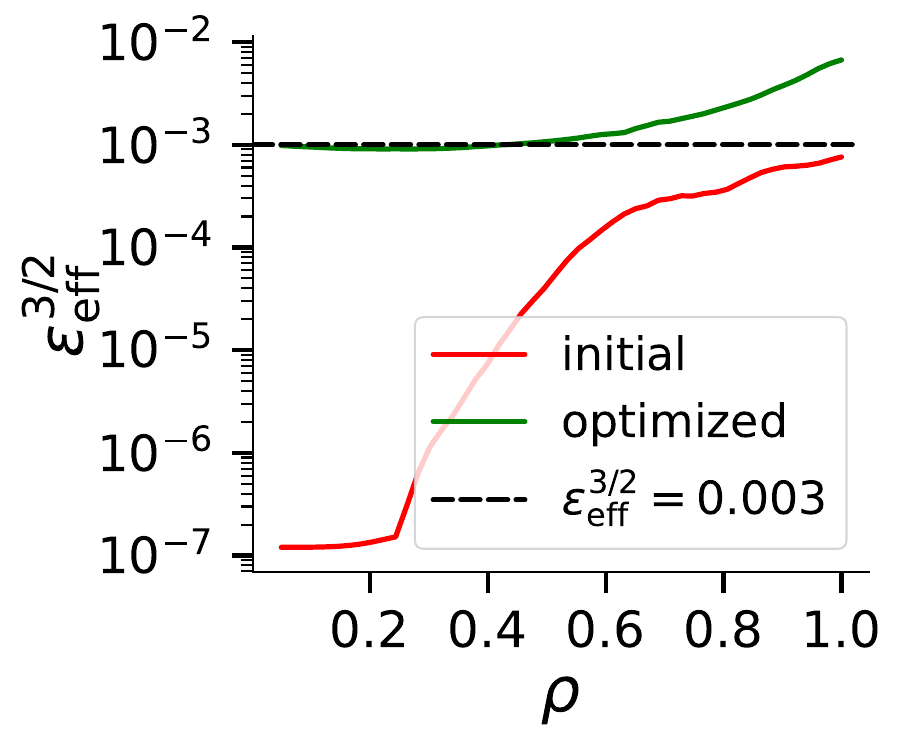}
        \caption{Neoclassical transport}
    \end{subfigure}
    \caption{Output from the vacuum US131 optimization showing Boozer plots on the boundary and the effective ripple. Figure~\textit{(b)} does not have a specific helicity but still achieves a low ripple transport.}
\label{fig:vacuum-result-Boozer}
\end{figure}

We begin by calculating an omnigenous vacuum equilibrium with a single field period $\NFP = 1$, $n=3, m=1$. To do this, we start with the US131 boundary shape, using the parametrization from section~\ref{subsec:UToL-parametrization} with $\hat{\rho}=0.1 \rm{m}, r_1 = 0.133, r_5 = 0.1$ and the rest of the parameters $r_j = 0$. The toroidal flux enclosed by the initial equilibrium $\psi_{\rm{b}} = \pi \hat{\rho}^2 (\rm{T} m^2)$. This shape is then approximated by~\texttt{DESC} as explained in section~\ref{subsec:fitting-UToLs} and optimized as described in section~\ref{subsec:optimizing-UToLs} with $w_{O} = 0$~\footnote{By setting $w_O = 0$ and $w_R \neq 0$, we seek omnigenous configurations without constraining the magnetic field to a specific helicity. This improves the quality of the optimum. This could also be a reason we are more likely to get piecewise-omnigenous configurations.} to obtain an omnigenous equilibrium with an aspect ratio $A = 10.5$. The results are presented in figures~\ref{fig:vacuum-result} and~\ref{fig:vacuum-result-Boozer}.

To ensure that the high-curvature umbilic edge is aligned with the magnetic field lines, the boundary undergoes deformation, which results in an increased rotational transform. The optimized design minimized the objective $\mathcal{F}_{\mathrm{stage-one}}$ while ensuring that the normalized second principal curvature is the lowest along the umbilic edge.  The curvature on the umbilic edge ranges from $\kappa_{2, \rho} \in [-187, -70]\, m^{-1}$ whereas the average curvature on the rest of the boundary is $\mathrm{avg}(\kappa_{2, \rho}) = -12 \,m^{-1}$. This causes the magnetic field line to stay close to the umbilic edge throughout the range $\zeta\in [0, 6 \pi]$ as shown in figure~\ref{fig:vacuum-result}\textit{(c)}. In the limit where the edge is perfectly sharp, the field line will coincide with the umbilic edge.

The optimized US case is able to achieve an effective ripple $\epsilon_{\mathrm{eff}}$ of less than $2\%$ ($\epsilon_{\mathrm{eff}}^{3/2} < 0.0028$) for most of the volume $(\rho < 0.86)$ and $\epsilon_{\mathrm{eff}}^{3/2} <= 0.001$ for $\rho <= 0.6$. This ensures that neoclassical transport in the $1/\nu$ low-collisionality regime will not deteriorate the device performance in the core but may deteriorate confinement in the edge. An interesting observation is that we can achieve a low ripple transport despite the lack of a specific helicity of the $B$ contours in the Boozer plot figure $4\mathit{(b)}$. This is an instance of a piecewise omnigenous~\citep{velasco2024piecewise} field that is far from quasisymmetry. Since we did not seek omnigenity with a specific helicity ($w_{\mathrm{O}} = 0$), the effective ripple penalty acted as a proxy for omnigenity, leading to a piecewise-omnigenous configuration. To prove that these configurations are piecewise omnigenous, we analyze the normalized second adiabatic invariant $\tilde{\mathcal{J}}_{\parallel}$ in Appendix~\ref{sec:pwO_plots}.

For a vacuum problem, obtaining the right fieldline pitch ($\iota = \NFP m/n$) on the boundary and ensuring force balance over a perfectly sharp (unapproximated) boundary makes this problem difficult to solve compared to a finite-$\beta$ equilibrium where one can specify the field line pitch throughout the volume. Therefore, we have compromised by creating a high-curvature edge instead of a perfectly sharp edge ($\kappa_{2, \rho} \rightarrow \infty$). If one were to use an alternate representation to represent a perfectly sharp boundary and solve this problem exactly, we would observe the rotational transform almost identical to the initial equilibrium in figure~\ref{fig:vacuum-result-Boozer}\textit{(c)} sharply rising to $\iota = m\, \NFP/n$ at the edge.The behavior of high-shear regions near sharp points would be similar to what is observed in tokamaks.


\subsection{Modular coil design for the US131 equilibrium}
\label{subsec:US131_coil}
\begin{figure}
    \centering
    \includegraphics[width=\textwidth, trim={1mm 1mm 1mm 15mm}, clip]{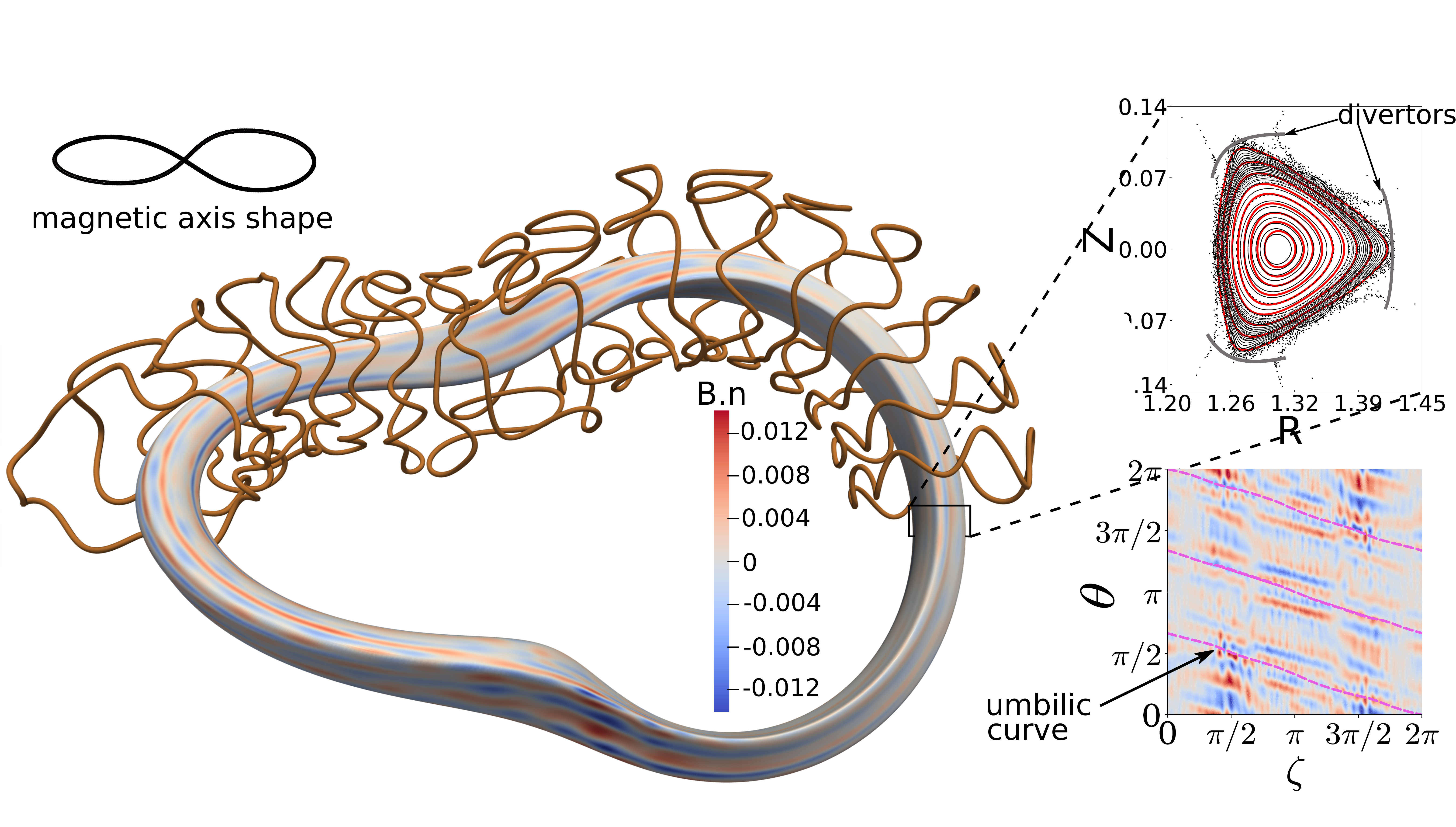}
    \caption{Coils and $\bm{B}_{\mathrm{out}}\cdot\hat{\bm{n}}$ error plotted on the plasma boundary. The $\mathrm{max}|(\bm{B}_{\mathrm{out}}\cdot\hat{\bm{n}})/\bm{B}_{\mathrm{out}}| \leq 1\%$. Only one group of unique coils are plotted. On the upper right we plot the Poincar\'{e} cross-section at $\zeta=0$ along with the fixed-boundary solution(in red). The gray curves mark a favorable position to place a divertor. In the lower right part we plot the $\bm{B}_{\mathrm{out}}\cdot\hat{\bm{n}}$ error on a 2D plane along with the umbilic edge(dashed purple).}
\label{fig:UT131-coils}
\end{figure}

In this section, we will design modular coils for the US131 case using the~\texttt{DESC} optimizer and elucidate the various difficulties faced when designing modular coils for strongly-shaped umbilic stellarators.
We create a coilset of $32$ (stellarator-symmetric with $16$ unique coils) coils using the Python implementation of \texttt{REGCOIL}~\citep{landreman2017improved, panici2025surface} in~\texttt{DESC}. Starting with this initial coilset, we then perform stage-two optimization, where we fix the equilibrium boundary and allow the shape of the modular coils to change while minimizing the objective 
\begin{eqnarray}
    \mathcal{F}_{\rm{stage-two}} = (B_{\rm{out}}^2 - (B_{\rm{in}}^2 + 2 \mu_0 p))^2 + (\bm{B}_{\rm{out}} \cdot \bm{\hat{n}})^2 
    \label{eqn:second-stage-penalty}
\end{eqnarray}
on the plasma boundary. Note that in stage-two optimization, we ignore the surface current $\bm{K} = (\bm{B}_{\rm{out}}   - \bm{B}_{\rm{in}}) \times \bm{\hat{n}}$. Therefore, to verify our solution, we calculate the field line trajectory due to the coils and create a Poincar\'{e} plot.  The original fixed-boundary solution, its comparison with the Poincar\'{e} cross-section, and the optimized coilset are presented in figure~\ref{fig:UT131-coils}.

Note that even though the curvature along the umbilic ridge is high, it does not behave like a perfect X-point (or X-line) as the fieldline structure becomes chaotic in the edge and resembles the ``turnstiles'' topology discussed by~\cite{punjabi2022magnetic}. However, even though the field lines do not exit in the regions of the highest curvature, like a tokamak X-point, the Poincar\'{e} traces of the fieldlines stay aligned with the plasma boundary and flare near the umbilic edge, which still makes the umbilic edge a favorable location to place divertors or limiters. This study indicates that the behavior of magnetic field lines for a high-curvature ridge can differ significantly from a perfectly sharp edge.

Another observation is that modular coil design becomes complicated for umbilic equilibria as seen in the figure. The coils assume high curvature to reduce the normal field error along the umbilic edge. Increasing the number of coils improves the agreement of the fixed boundary solution with the Poincar\'{e} cross-section and reduces the max coil curvature but also reduces the coil-coil distance. However, the number of modular coils needed for these umbilic equilibria already exceeds the number of coils used for a single field period for a typical stellarator, which is almost always less than ten. 

Finally, we note that the regions of maximum normal field error are parallel to the umbilic edge so that, on average, $\bm{b} \cdot \bm{\nabla} (\bm{B}_{\mathrm{out}}\cdot \hat{\bm{n}}) = 0$\footnote{We use the magnetic field unit vector instead of the vector along the umbilic edge because the magnetic field is reasonably aligned with the umbilic edge throughout its span as shown in figure~\ref{fig:vacuum-result}\textit{(c)}}. This pattern indicates a hidden relationship between the normal field error and the second principal curvature in umbilic stellarators.

To ensure practical coil design for umbilic stellarators, a better technique would be to perform single-stage optimization, which simultaneously changes the plasma boundary and coil shapes, a union of $\mathcal{F}_{\mathrm{stage-one}}$ and $\mathcal{F}_{\mathrm{stage-two}}$, accounting for the coil complexity while maintaining omnigenity inside the plasma volume. This is done in~\S\ref{subsec:co-I} for a simpler problem. Another possibility is to use umbilic coils, helically wound divertor-like coils with a fractional helicity($m/n$) that carry a small current that pushes (pulls) the plasma away from (towards) itself, creating cusp-like regions, leading to a high curvature umbilic edge. We do a similar optimization to modify a tokamak in section~\S\ref{subsec:counter-I}, where we add an umbilic coil without changing the shape of the toroidal field coils to create an umbilic edge. In the next section, we shall use the same optimization process to create a finite-$\beta$, omnigenous umbilic stellarator.

\subsection{US131 sensitivity analysis of the field structure in the edge}
\begin{figure}
    \centering
    \begin{subfigure}[b]{0.31\textwidth}
    \centering
        \includegraphics[width=\textwidth, trim={0mm 2mm 2mm 2mm}, clip]{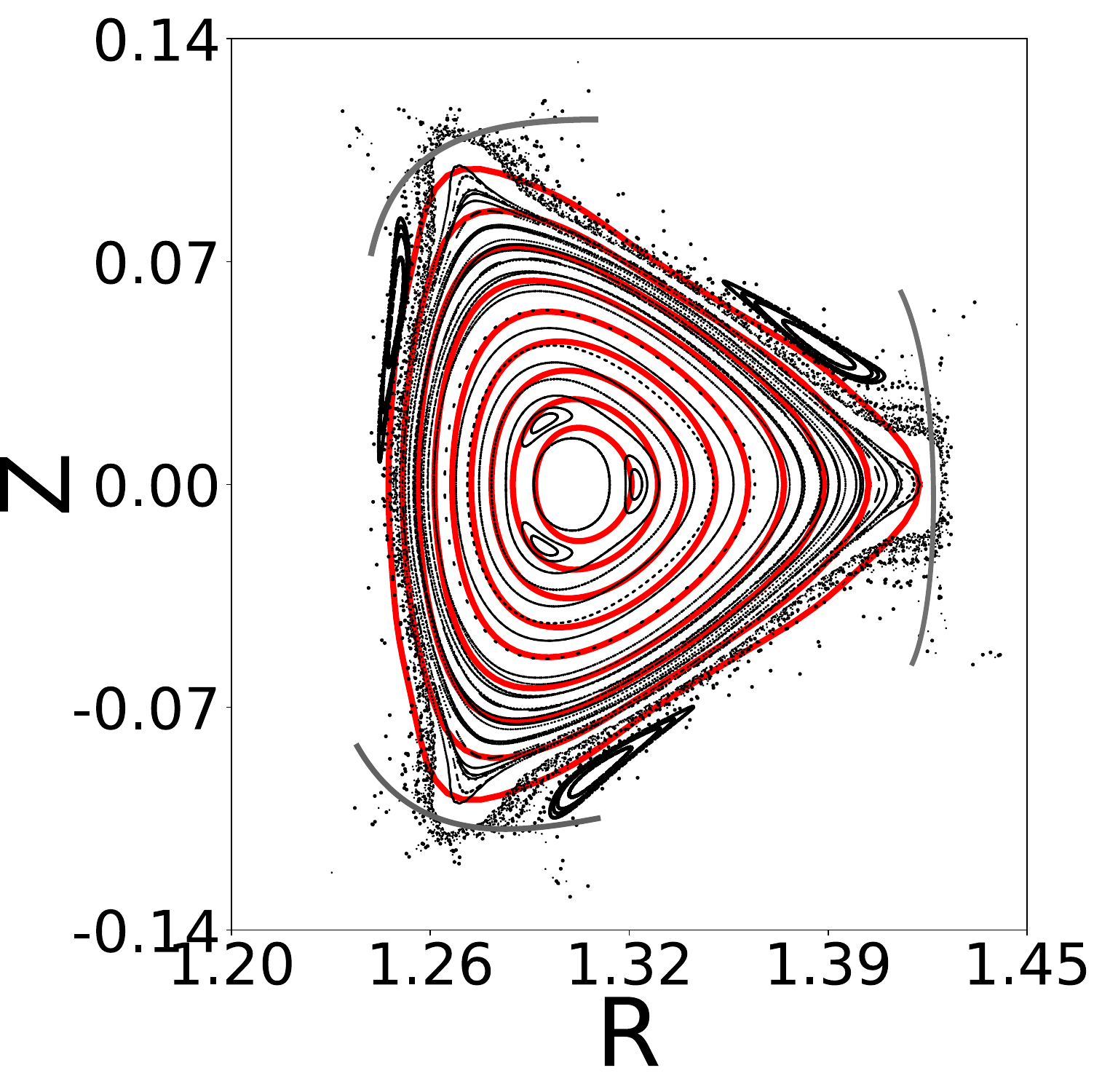}
        \caption{$I_{\mathrm{axis}}=250 A$}
    \end{subfigure}
    \hspace*{-1mm}
    \begin{subfigure}[b]{0.31\textwidth}
    \centering
        \includegraphics[width=1.0\textwidth, trim={0mm 2mm 2mm 2mm}, clip]{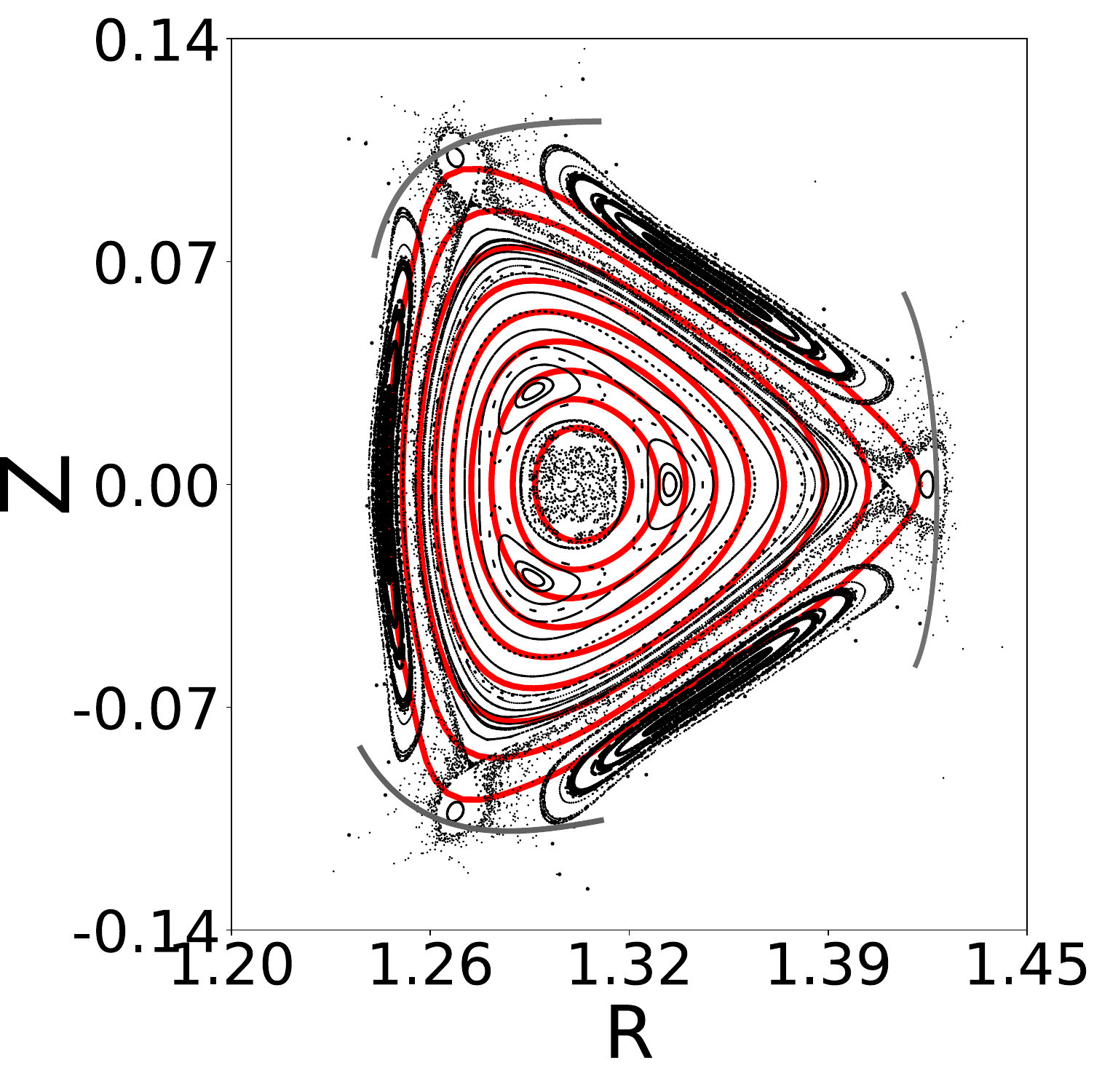}
        \caption{$I_{\mathrm{axis}}=500 A$}
    \end{subfigure}
    \hspace*{-1mm}
    \begin{subfigure}[b]{0.31\textwidth}
        \captionsetup{margin={1mm,0cm}}
        \centering
        \includegraphics[width=\textwidth, trim={0mm 2mm 2mm 2mm}, clip]{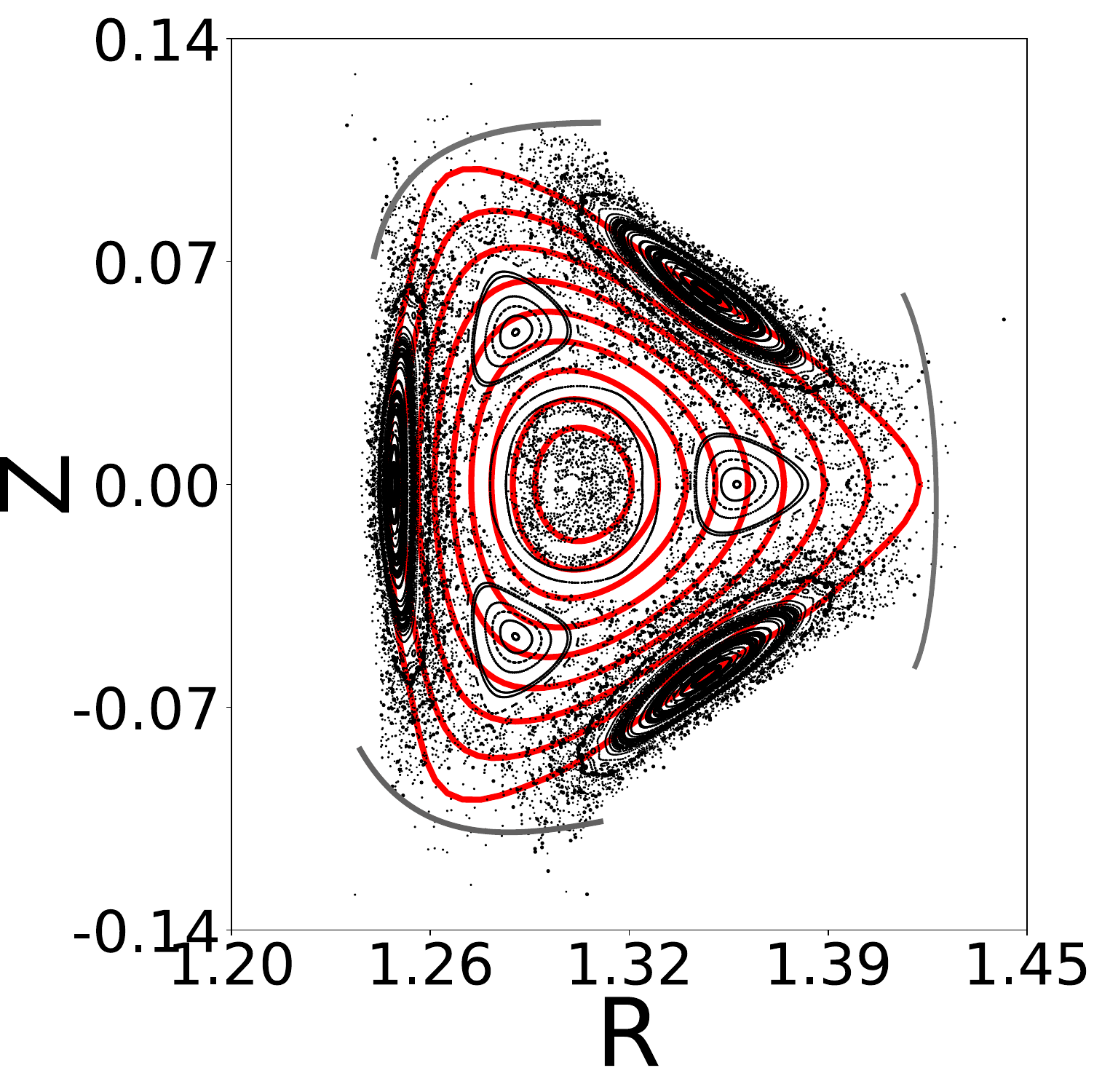}
        \caption{$I_{\mathrm{axis}}=1000 A$}
    \end{subfigure}
    \caption{Poincare plots at $\zeta = 0$ with a current source placed on-axis for different values of the current, compared with the original fixed boundary solution (in red). Increasing the current significantly reduces the nestedness of the flux surfaces but the fieldline structure in the edge still ``flares'' out near the high-curvature edges and the divertor placement does not need to be modified.}
\label{fig:US131-perturbation}
\end{figure}

In this section, we shall introduce a line current source along the magnetic axis of the US131 configuration. The line current aims to simulate an internal fluctuation within the plasma. For different values of the current source strength, we will perform the field line integration across the entire volume. The purpose of this investigation is to assess the resilience of the field line structure at the edge of the US131 stellarator and to determine the extent of modification required for the divertor configuration. 
We present the fieldline structure for different values of the current in figure~\ref{fig:US131-perturbation}.

We observe that even though the flux surfaces inside the plasma boundary lose nestedness and become stochastic, the divertor structure used for the unperturbed US131 case is still an effective choice for this equilibrium, i.e., the fieldline structure in the edge is resilient to fluctuations inside the plasma. Another observation is the emergence of ``O''-point-like structures near the high-curvature regions in figure~\ref{fig:US131-perturbation}\textit{(b)}. 

\subsection{US252 $(\NFP=2, n=5, m=2)$ omnigenous finite-$\beta$ stellarator}
\label{subsec:UT135}
In this section, we will generate a two-field period $n=5, m = 2$ finite-$\beta$ stellarator. We start with the umbilic parametrization defined in subsection~\ref{subsec:UToL-parametrization} with $\varrho=0.1, r_1 = 0.133, r_5 = 0.1$ and use a pressure profile $p = 5 \times 10^3 (1-\rho^2) \mathrm{Pa}$, a rotational transform $\iota = 0.755 + 0.043 \rho^2$, and an enclosed toroidal flux $\psi_{\rm{b}} = \pi \hat{\rho}^2 = 0.00314\, \rm{T} m^2$. This creates an equilibrium with a $\beta = 0.5\%$.

\begin{figure}
    \centering
    \begin{subfigure}[b]{0.33\textwidth}
    \centering
        \includegraphics[width=\textwidth, trim={1mm 1mm 1mm 2mm}, clip]{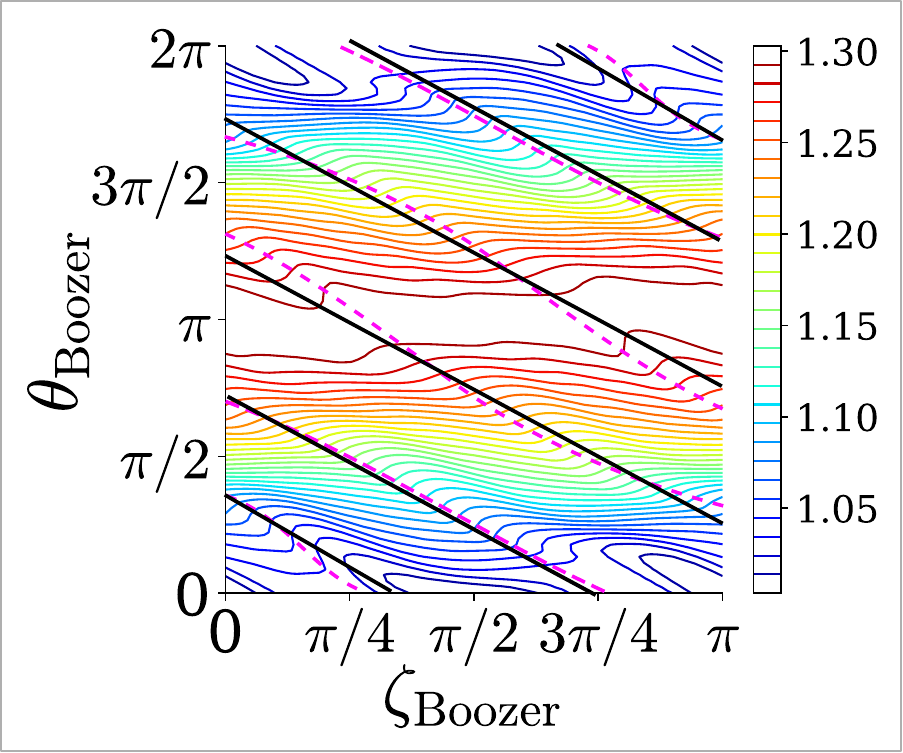}
        \caption{Initial $|B|(\rho = 1.0)$}
    \end{subfigure}
    \hspace*{-1mm}
    \begin{subfigure}[b]{0.325\textwidth}
    \centering
        \includegraphics[width=1.0\textwidth, trim={2mm 2mm 2mm 2mm}, clip]{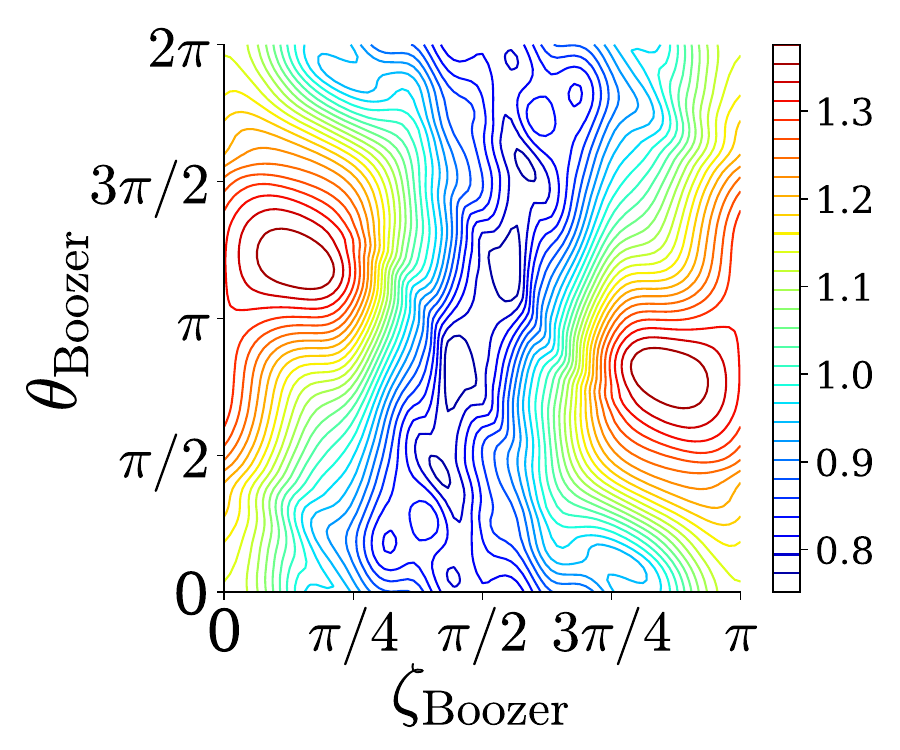}
        \caption{Optimized $|B|(\rho = 1.0)$}
    \end{subfigure}
    \hspace*{-1mm}
    \begin{subfigure}[b]{0.33\textwidth}
        \captionsetup{margin={1mm,0cm}}
        \centering
        \includegraphics[width=\textwidth, trim={0mm 1mm 0mm 2mm}, clip]{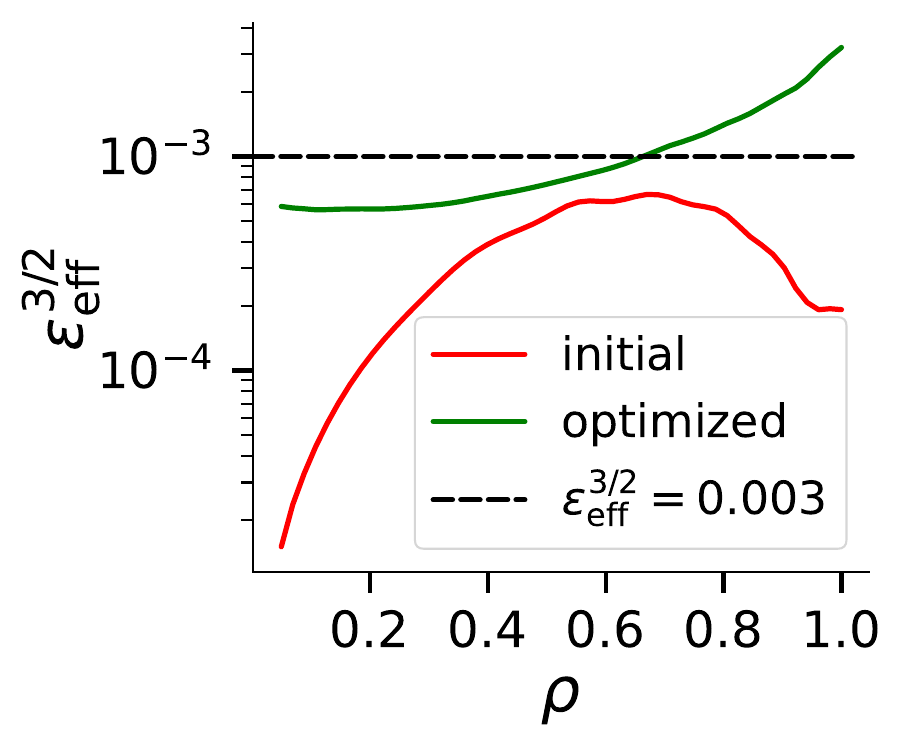}
        \caption{Neoclassical transport}
    \end{subfigure}
    \caption{Optimization output showing the initial and final magnetic field strength Boozer plots in figures \textit{(a)} and \textit{(b)}, respectively and the effective ripple in figure $\textit{(c)}$. The black lines \textit{(a)} correpond to the magnetic field lines and the dashed lines correspond to the umbilic curve. in The optimized configuration is another instance of piecewise omnigenity}
\label{fig:UT225-result}
\end{figure}

 A finite-$\beta$ stellarator provides more freedom compared to a vacuum case because the rotational transform can come from a combination of boundary shaping and plasma current. Therefore, we can achieve a higher rotational transform and use the additional freedom to achieve a much lower ripple transport compared to the vacuum case presented in the previous section. The optimization uses the same objective function defined in~\eqref{eqn:overall-objective} with the pressure and iota profiles fixed and $w_{\iota} = 0$. These results are presented in figure~\ref{fig:UT225-result}. 

We can see in figure~\ref{fig:UT225-result}, the ridges in the Boozer plots are only clearly visible in the initial equilibrium, which is close to quasisymmetry. The magnetic field of the optimized omnigenous equilibrium does not have any visible ridges in the Boozer plot, even though the curvature along the umbilic ridge increases after optimization. This indicates that the appearance of ridges in the Boozer representation may only be a feature of quasisymmetric stellarators.

\begin{figure}
    \centering
    \begin{subfigure}[b]{0.42\textwidth}
    \centering
        \includegraphics[width=\textwidth, trim={0mm 1mm 0mm 2mm}, clip]{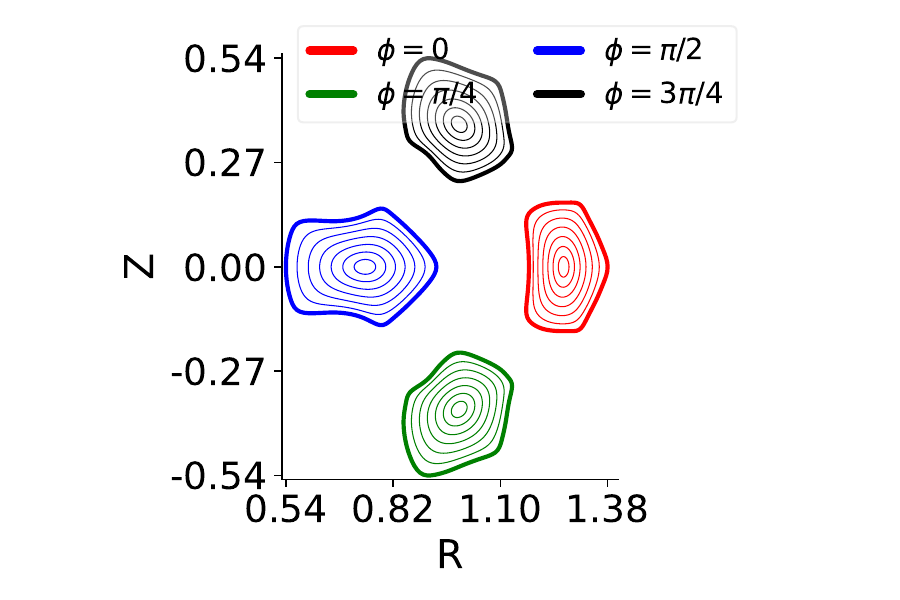}
        \caption{Optimized cross-sections}
    \end{subfigure}
    \hspace*{-10mm}
    \begin{subfigure}[b]{0.32\textwidth}
    \centering
        \includegraphics[width=\textwidth, trim={0mm 20mm 0mm 10mm}, clip]{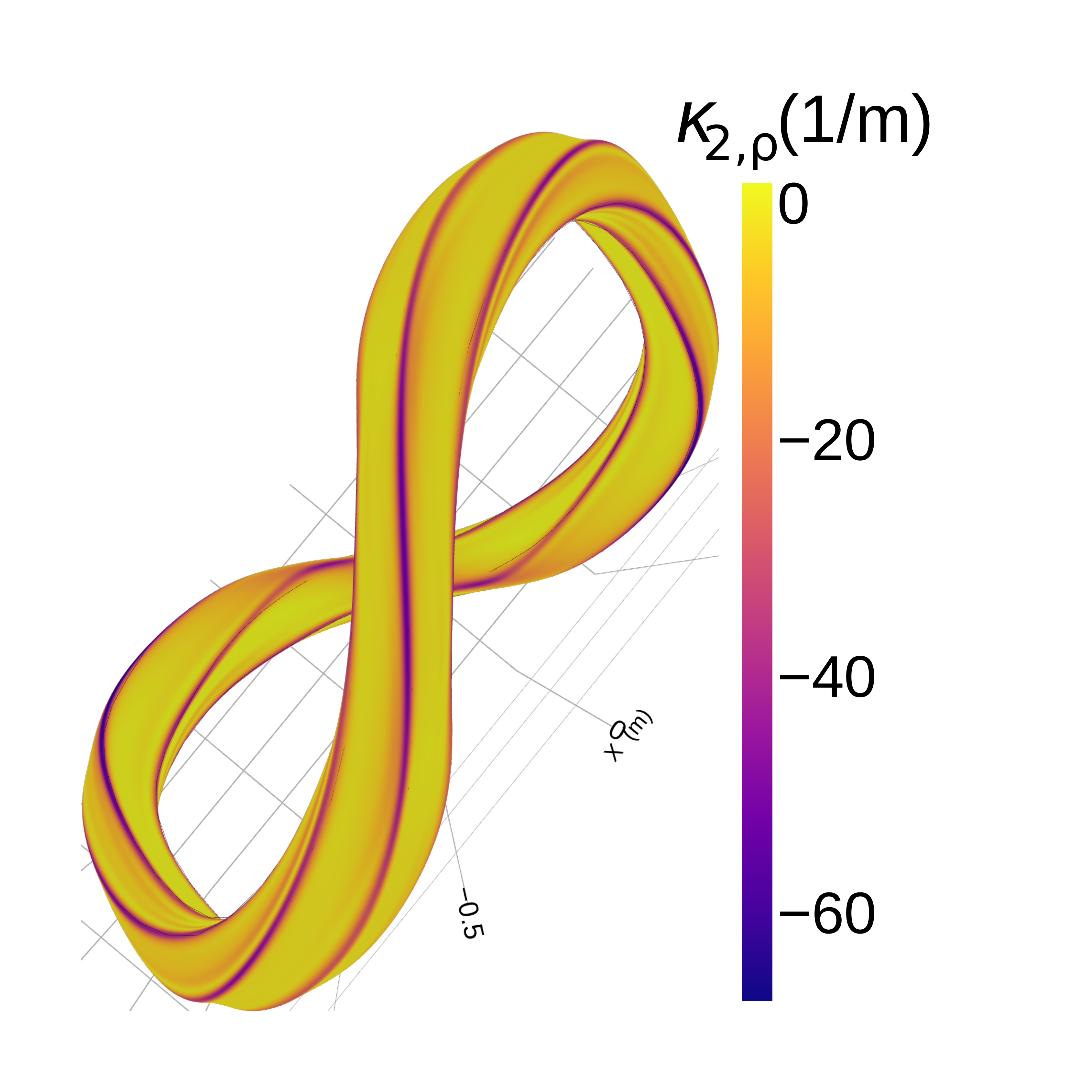}
        \caption{Curvature on boundary}
    \end{subfigure}
    \hspace*{-5mm}
    \begin{subfigure}[b]{0.30\textwidth}
    \centering
        \includegraphics[width=\textwidth, trim={0mm 2mm 1mm 2mm}, clip]{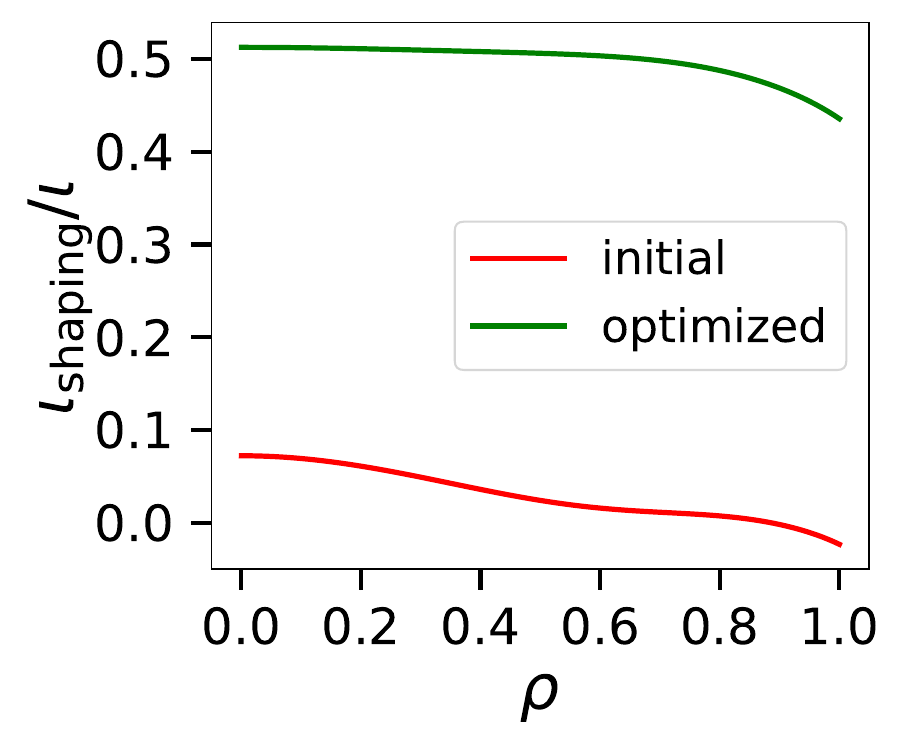}
        \caption{$\iota$ fraction from shaping}
    \end{subfigure}
    \caption[Caption for LOF]{Cross-section and three-dimensional shape of the optimized US252 equilibrium. The aspect ratio of the optimized equilibrium is $A = 6.5$. In figure~\textit{(c)}, we see that around half of the rotational transform of the optimized equilibrium comes from shaping~\citep{HIRSHMAN1986329} reducing the dependence on external current drive.}
    \label{fig:UT225-result2}
\end{figure}

This equilibrium is optimized for poloidal omnigenity but, due to the umbilic curvature penalty, limits the poloidal omnigenity to the low-field region. The Boozer plot in figure~\ref{fig:UT225-result}\textit{(b)} looks qualitatively like a piecewise-omnigenous~\cite{velasco2024piecewise} configuration, which looks similar to the magnetic field in the W7-X stellarator. The piecewise omnigenity of this equilibrium is demonstrated further in Appendix~\ref{sec:pwO_plots}. We also plot the cross-section of the equilibrium along with the second principal curvature in figure~\ref{fig:UT225-result2}. The neoclassical transport coefficient $\epsilon_{\mathrm{eff}}^{3/2} < 0.001$ for most ($\rho < 0.7$) of the volume, which means the equilibrium is omnigenous in the core.

We also observe that the optimized equilibrium deforms and rotates as we move toroidally, while maintaining high curvature along the umbilic curve. The curvature along the umbilic edge $\kappa_{2, \rho} \in [-67, -36]\, m^{-1}$ while the average curvature on the rest of the plasma boundary is $\mathrm{avg} (\kappa_{2, \rho}) = -8\, m^{-1}$. The variation in the curvature is broad, indicating that imposing high curvature at specific locations may be more suitable than others. This can be readily accomplished by selecting sections of the umbilic curve where high curvature is applied and could be a potential way to place potential locations for non-resonant divertors~\citep{bader2018minimum}.

\section{Umbilic coil design for a finite-$\beta$ equilibrium}
\label{sec:HBT-coil}
As we have seen in section~\ref{subsec:vacuum-UToL}, creating modular coils for vacuum umbilic equilibria is a difficult task due to the sharp curvature regions. Therefore, in this section, we present an alternate strategy that uses umbilic coils to create a US111 finite-$\beta$ equilibrium. Instead of modifying modular coils, we shall use an umbilic coil, a low-current, helically wound coil, similar to a divertor coil in tokamaks, which can be used to create an umbilic edge. Unlike tokamak divertor coils, an umbilic coil is non-axisymmetric and can have a fractional helicity. In the following sections, we will discuss two cases, categorized by the direction of the current in the umbilic coils with respect to the toroidal plasma current.

\subsection{US111 with an umbilic coil current opposite to the plasma current (counter-$I$ case)}
\label{subsec:counter-I}
Umbilic coils differ from helical coils by creating a ridge that may serve as a potential location for an X-point. These coils are similar to tokamak divertor coils, albeit without axisymmetry. To demonstrate their applicability, we modify a $n=1$ MHD-stable equilibrium from the High-Beta Torus-Extended Pulse (HBT-EP) experiment ($R_0 = 0.94 m, A = 7.5$) by adding an $m = 1, n = 1$ ridge to it, breaking its axisymmetry, and then design an umbilic coil for it without modifying the toroidal or vertical field coils. The specific details of the equilibrium used are given in Appendix~\ref{app:HBT-EP}.

\begin{figure}
    \centering
    \begin{subfigure}[b]{0.36\textwidth}
    \centering
        \includegraphics[width=\textwidth, trim={0mm 20mm 0mm 2mm}, clip]{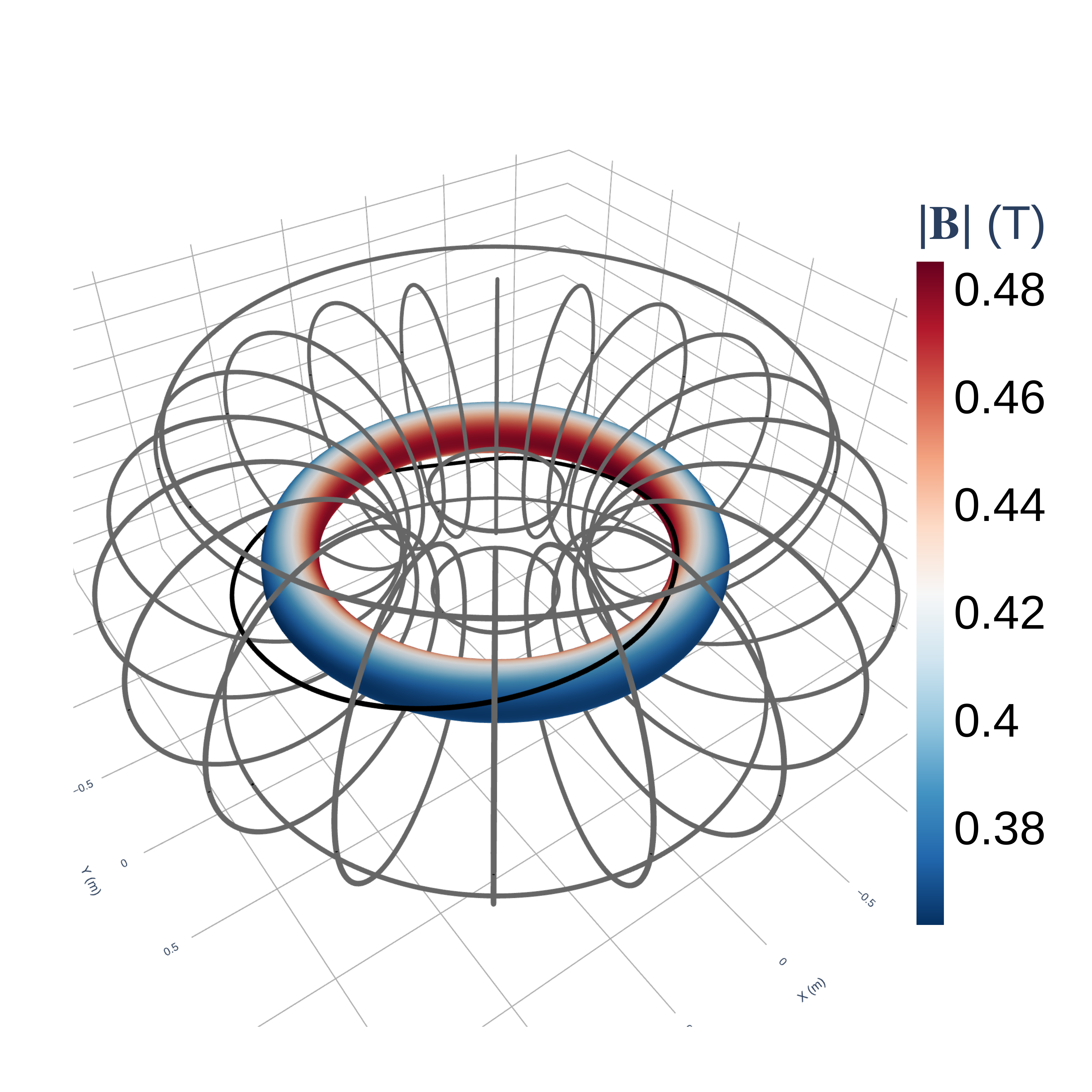}
        \caption{Mod. HBT-EP full coilset}
    \end{subfigure}
    \hspace*{-1mm}
    \begin{subfigure}[b]{0.33\textwidth}
        \captionsetup{margin={2mm,0cm}}
        \centering
        \includegraphics[width=1.05\textwidth, trim={5mm 20mm 20mm 2mm}, clip]{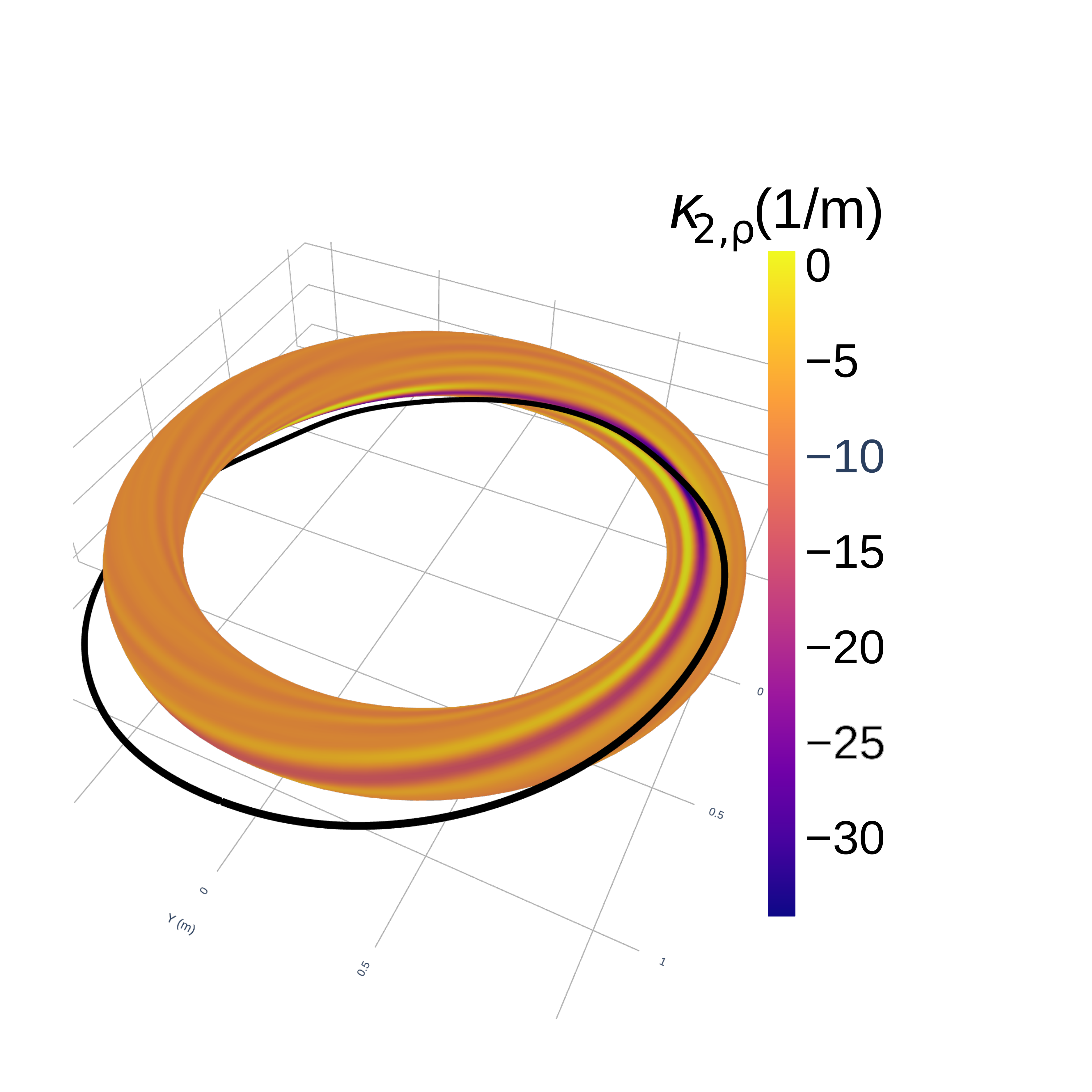}
        \caption{$\kappa_{2, \rho}$ \& umbilic coil}
    \end{subfigure}
    \hspace*{-7mm}
    \begin{subfigure}[b]{0.30\textwidth}
        \captionsetup{margin={0mm,0cm}}
        \centering
        \includegraphics[width=0.9\textwidth, trim={0mm -8mm 0mm 2mm}, clip]{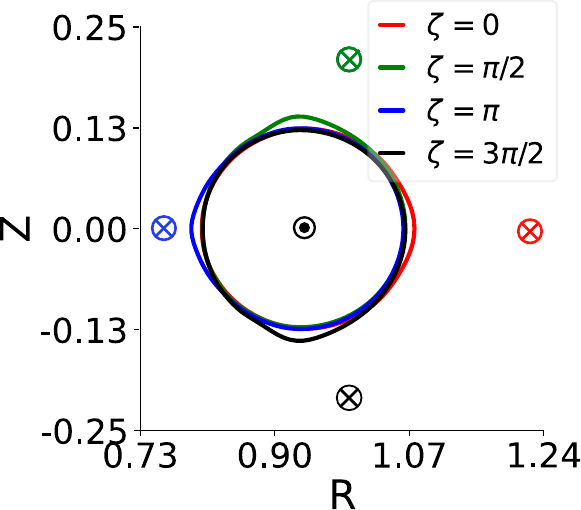}\\[3mm]
        \caption{X-section \& umbilic coil}
    \end{subfigure}
    \caption{Output from the stage-two optimization and free-boundary solve. Figure~\textit{(a)} shows the full coilset of the modified HBT-EP experiment with the magnetic field strength on the boundary whereas figure~\textit{(b)} shows the umbilic coil and the curvature $\kappa_{2, \rho}$ on the boundary. Figure~\textit{(c)} shows the cross section of the LCFS from the free-boundary \texttt{DESC} solver along with the umbilic coil at four different toroidal angles and the directions in which they carry current relative to the plasma.}
\label{fig:HBT-EP-result}
\end{figure}

To accomplish this, we start by using $20$ Toroidal Field (TF)~\footnote{In the HBT-EP experiment the TF coils have been wound multiple times and packed into a square Aluminum frame with finite thickness of over $0.15 \rm{m}$ whereas we assume filamentary TF coils in our calculations, ignoring the effect of finite thickness. The radius of the TF filamentary coils is $0.65 \rm{m}$, which is the average of the outer and inner dimensions of a real HBT-EP TF coil.} coils and $4$ Vertical Field (VF) poloidal coils as done in the original HBT-EP experiment~\citep{gates1993passive, sankar1993initial} and add a $m = 1, n =1$ umbilic coil outside the plasma boundary at a distance of $1.5$ times the minor radius. Similar to section~\ref{subsec:US131_coil}, we perform second stage optimization to find the right shape of the umbilic coil. To ensure that the coils from stage-two optimization give expected results, we calculate the free-boundary solution and present it in figure~\ref{fig:HBT-EP-result}.

\begin{figure}
    \centering
    \includegraphics[width=0.9\textwidth, trim={0mm 0mm 0mm 0mm}, clip]{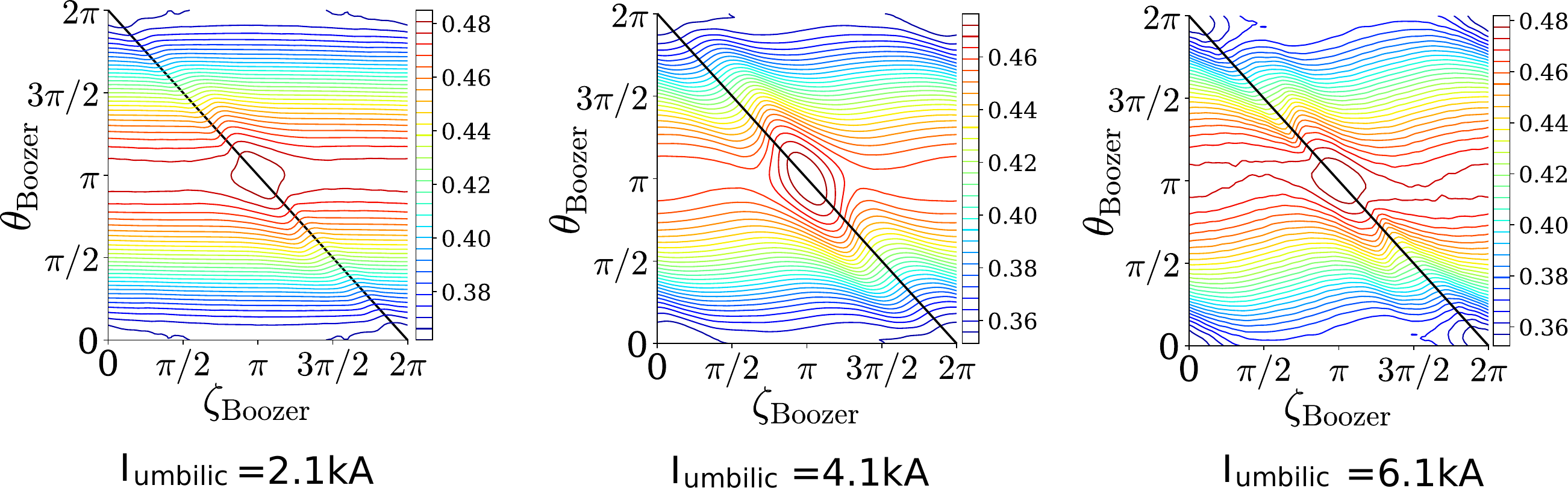}
    \caption{Magnetic field strength in Boozer coordinate on the boundary $(\rho=1.0)$ for different values of the umbilic coil current and the fieldline $\alpha = 0$. As the curvature of the ridge increases, so does the localized distortion of the $|\bm{B}|$ contours close to the fieldline.}
\label{fig:HBT-Boozer-plots}
\end{figure}

The effect of the umbilic coil can be decomposed as a rigid shift of the whole equilibrium and a small deformation of the boundary. Since the rigid shift~\citep{Sengupta_Nikulsin_Gaur_Bhattacharjee_2024} and the size of the ridge is small compared to the minor radius, it does not compromise omnigenity (as seen later in figure~\ref{fig:HBT-Boozer-plots}\textit{(a)}).
For a perfectly sharp umbilic edge, the magnetic field line must coincide with the edge, but even when the ridge lacks perfect sharpness, the magnetic field line stays close to it due to its high curvature. The curvature along the umbilic ridge $\kappa_{1, \rho} \in [-15, -32] m^{-1}$ whereas the average curvature $\mathrm{avg}(\kappa_{2, \rho}) = -7 m^{-1}$.

The TF coils each carry a current of $105\,\text{kA}$, the PF coils carry $29.8\,\text{kA}$, and the umbilic coil has a current $I_{\mathrm{umbilic}} = 2.1\,\text{kA}$. But due to the proximity of the umbilic coil compared to TF and PF coils, it can still affect the plasma near the boundary significantly. As we can see, the umbilic coil is able to create a $m=1, n=1$ high curvature ridge. The ridge formation is entirely attributed to the umbilic coil, as all other coils are axisymmetric. The current flowing through the coil is in the opposite direction to the plasma current. However, due to the 3D nature of this setup, the plasma still bulges toward the coil.

\begin{figure}
    \centering
    \begin{subfigure}[b]{0.325\textwidth}
        \captionsetup{margin={-8mm,0cm}}
    \centering
        \includegraphics[width=1.0\textwidth, trim={0mm 0mm 0mm 0mm}, clip]{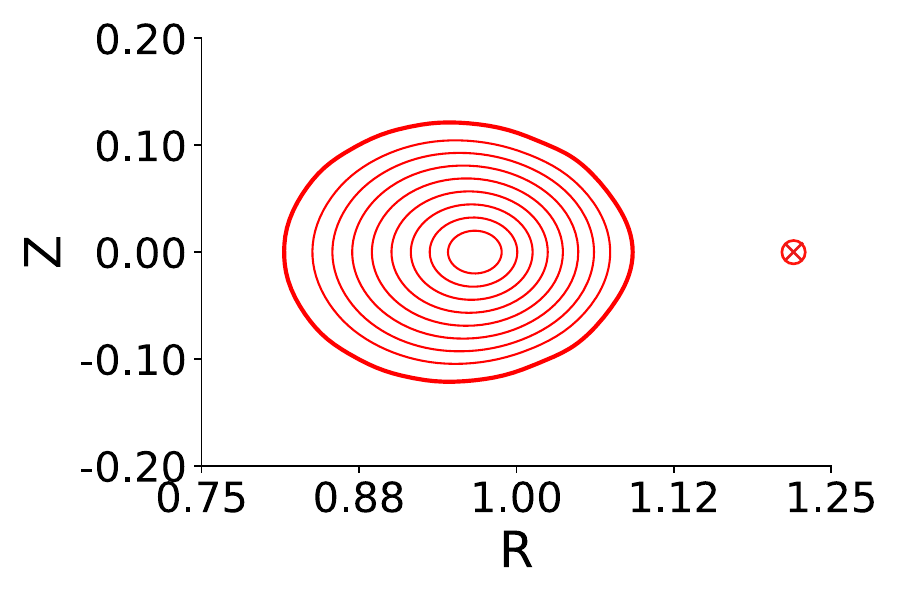}
        \caption{$I_{\mathrm{umbilic}}= 4.1 kA$}
    \end{subfigure}
    \begin{subfigure}[b]{0.31\textwidth}
        \captionsetup{margin={0mm,0cm}}
        \centering
        \includegraphics[width=1.08\textwidth, trim={0mm 0mm 0mm 0mm}, clip]{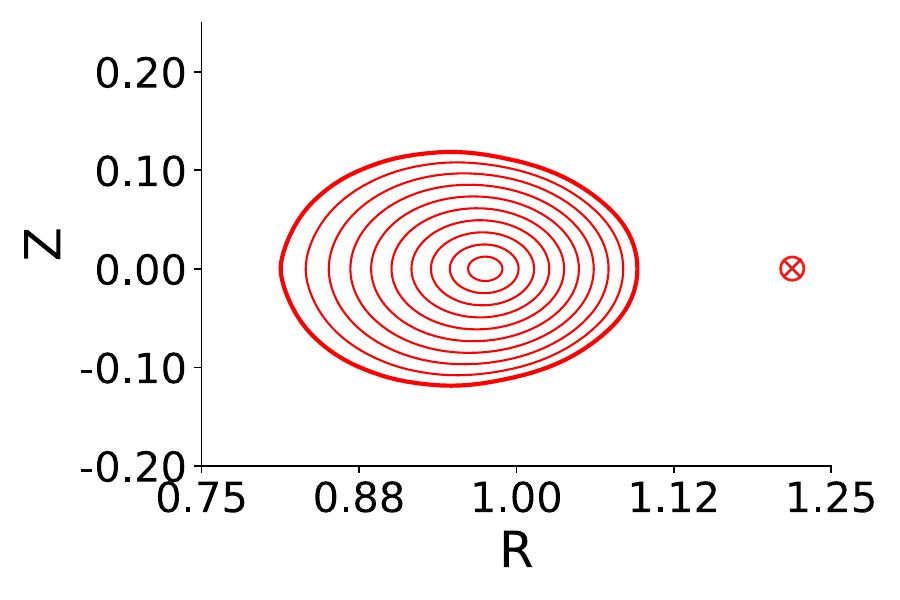}
        \caption{$I_{\mathrm{umbilic}}= 6.1 kA$}
    \end{subfigure}
    \begin{subfigure}[b]{0.3\textwidth}
        \centering
        \includegraphics[width=0.95\textwidth, trim={0mm 2mm 0mm 2mm}, clip]{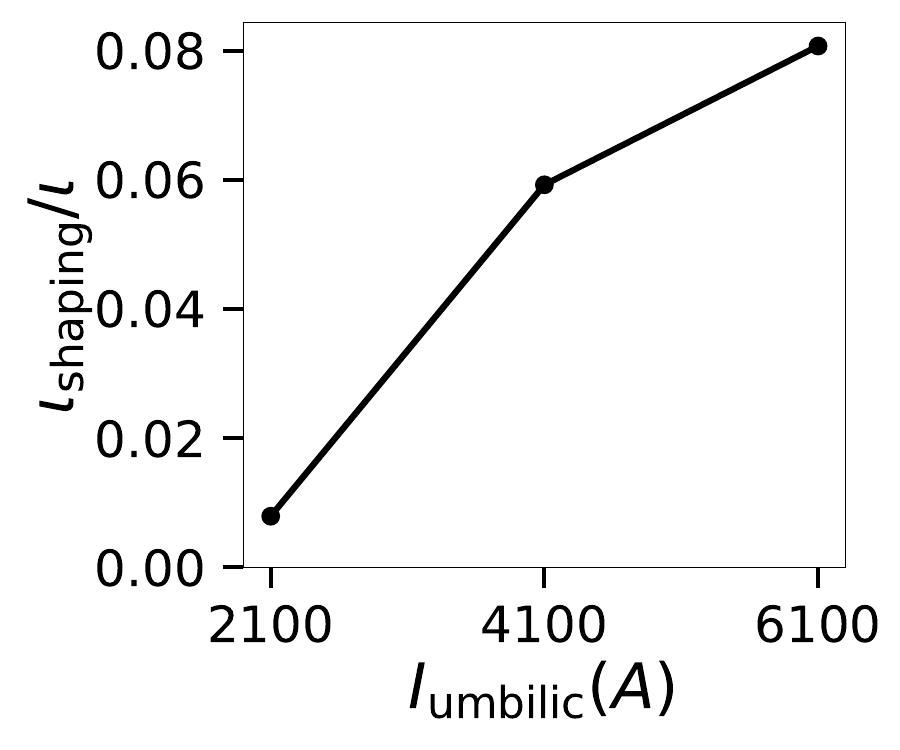}
        \caption{$\iota$ fraction from shaping}
    \end{subfigure}
    \caption{Free-boundary solution at $\zeta = 0$ for different values of the umbilic coil current $I_{\mathrm{umbilic}}$ shown in~\textit{(a)-(b)} and the fraction of the rotational transform generated by the umbilic coil current\protect\footnotemark in~\textit{(c)}.}
    \label{fig:FIELDLINES-high-current}
\end{figure}

To understand the structure of the magnetic field near the edge, we use the~\texttt{FIELDLINES}~ \citep{lazerson2016first} code from the~\texttt{STELLOPT}~\citep{doecode_12551} optimizer.
\begin{figure}
    \centering
    \begin{subfigure}[b]{0.3\textwidth}
    \centering
        \includegraphics[width=1.1\textwidth, trim={0mm 2mm 10mm 2mm}, clip]{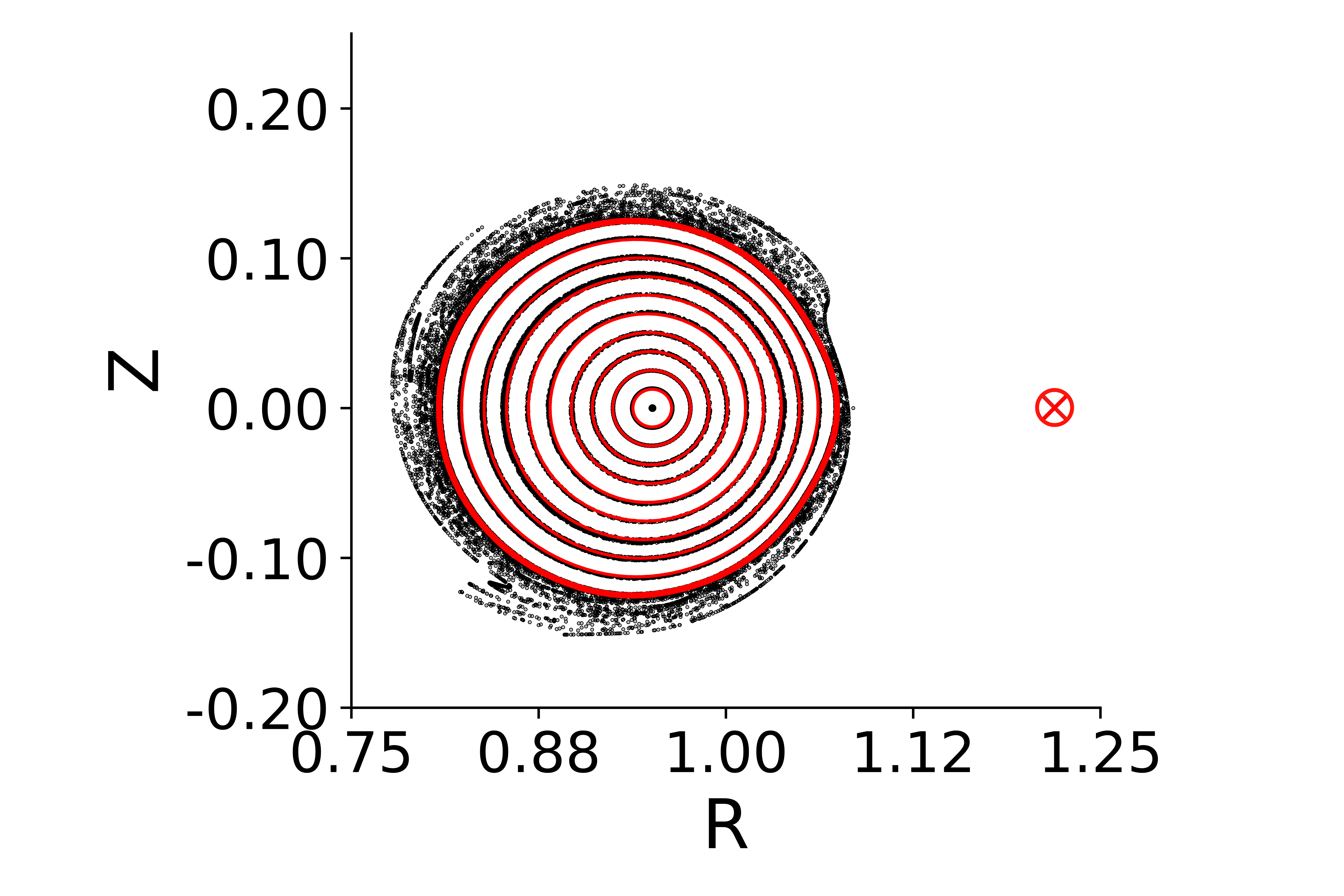}
        \caption{Poincar\'{e} section ($\zeta =0$)}
    \end{subfigure}
    \hspace*{-1mm}
    \begin{subfigure}[b]{0.34\textwidth}
        \captionsetup{margin={-0mm,0cm}}
        \centering
        \includegraphics[width=1.02\textwidth, trim={0mm 2mm 2mm 2mm}, clip]{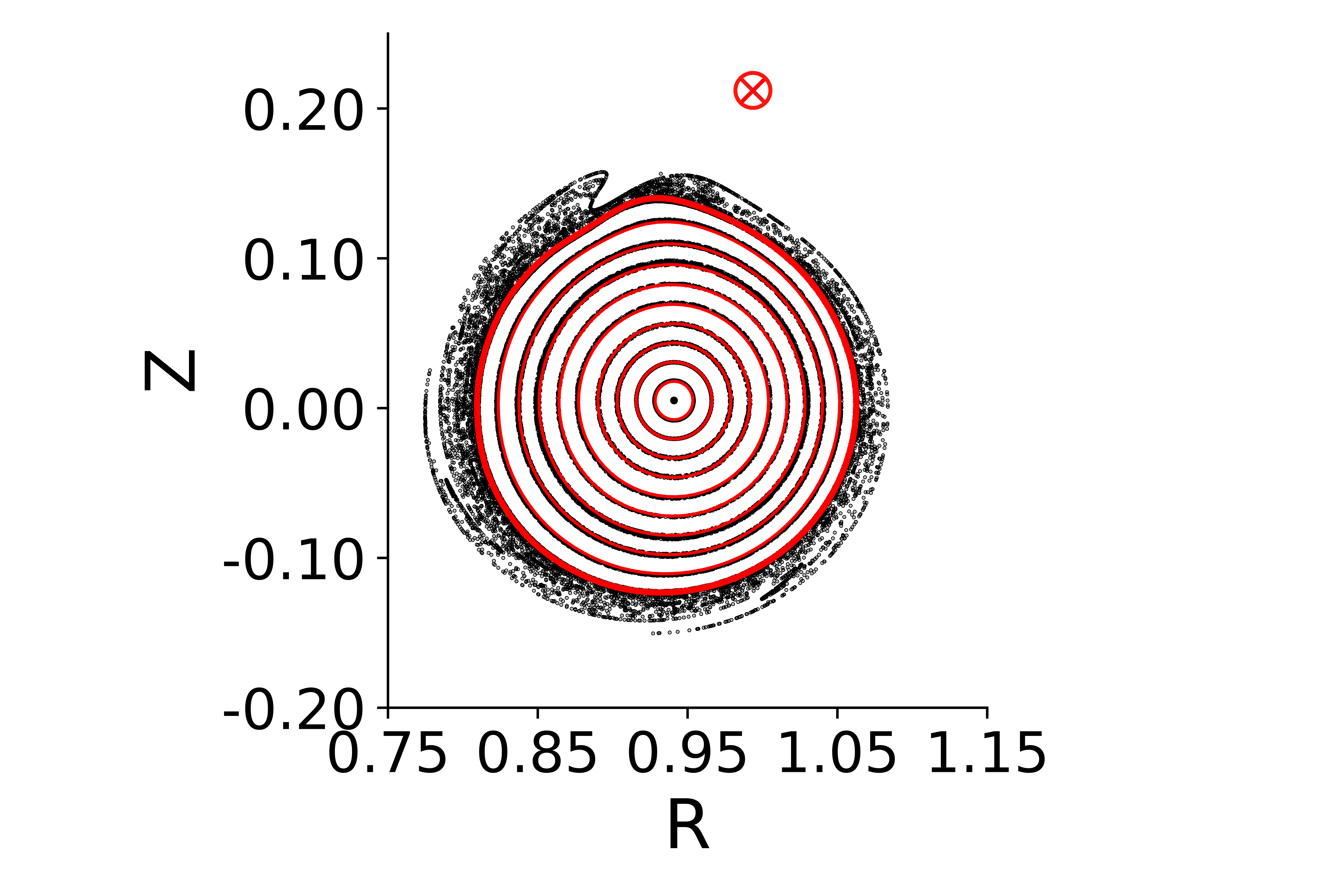}\\[-0.2mm]
        \caption{Poincar\'{e} section ($\zeta =\pi/2$)}
    \end{subfigure}
    \begin{subfigure}[b]{0.3\textwidth}
        \centering
        \includegraphics[width=0.8\textwidth, trim={0mm 0mm 0mm 2mm}, clip]{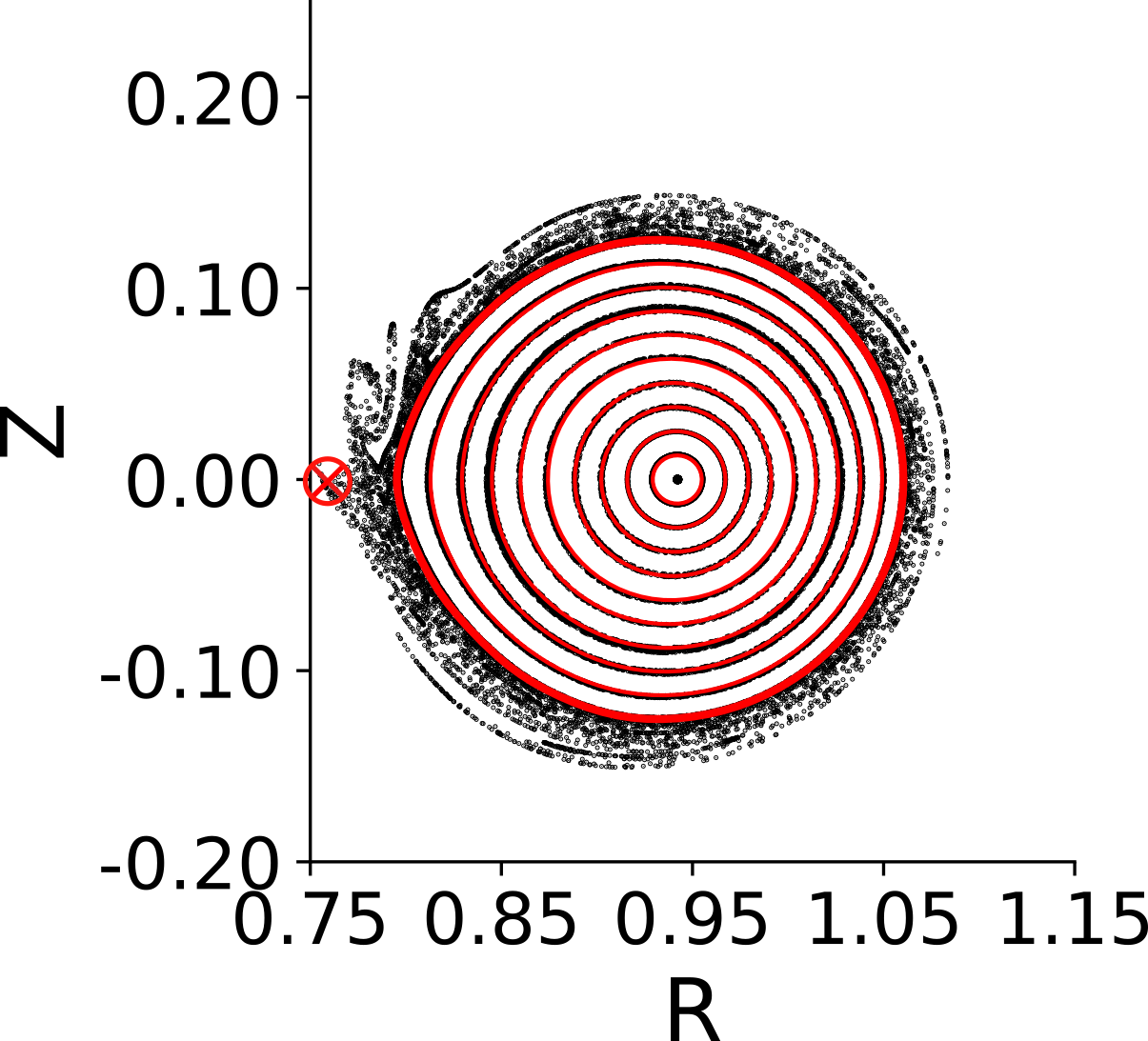}\\[0.5mm]
        \caption{Poincar\'{e} section ($\zeta =\pi$)}
    \end{subfigure}
    \caption{Poincar\'{e} section at different toroidal angles compared with free-boundary~\texttt{DESC} solution(red). The red cross marks the position of the umbilic coil.}
\end{figure}
We observe the presence of a stochastic layer outside the plasma boundary, similar to heliotron/torsatron configurations. Since the plasma and the umbilic coil are carrying currents in opposite directions, the poloidal field cannot vanish between them, which is why we do not observe an X-point. Therefore, the umbilic coil can create and vary the curvature of the ridge without creating and varying the current in the umbilic coil without creating an X point. This also affects the fraction of the rotational transform generated by the umbilic coils. These details are provided in figure~\ref{fig:FIELDLINES-high-current}.

\footnotetext{Since all the coils except the umbilic coil are axisymmetric and circular, any departure from a circular plasma shape is a consequence of the umbilic coil. Furthermore, for a perfectly circular equilibrium ($\iota_{\mathrm{shaping}} =0$), so all the rotational transform on the boundary must come from non-circular shaping. Therefore, we conclude that any finite $\iota_{\mathrm{shaping}}$ a consequence of the umbilic coil.}
By controlling the current in the umbilic coil, we can shape the plasma without moving the limiter. Also, the ability to vary $\iota$ would help to stabilize various MHD instabilities and the $m=1, n=1$ tearing mode as done in the RFX-mod experiment~\citep{piron2016interaction} or experiments like the Madison Spherical Torus, where the edge rotational transform close to one can be achieved~\citep{hurst2022self}. The J-TEXT~\citep{liang2019overview} tokamak has recently proposed a similar idea~\citep{LI2024114591} of using umbilic coils to stabilize the edge of a tokamak. 

Upon breaking the axisymmetry of the magnetic field, one must be careful as increasing the current $I_{\mathrm{umbilic}}$ for the same coil shape would degrade omnigenity and hence neoclassical transport. To understand the effect of umbilic coils on omnigenity, we present Boozer plots for the three values of current in figure~\ref{fig:HBT-Boozer-plots}.

We can see that the omnigenity degrades as we increase $I_{\mathrm{umbilic}}$ but the neoclassical losses are still acceptable $\epsilon_{\mathrm{eff}}^{3/2} < 0.001$. To control instabilities and transport due to MHD modes, we may need to generate a higher $\iota_{\mathrm{shaping}}$ while ensuring omnigenity. One way to explore multiple such coil and equilibrium configurations would be to perform a single-stage optimization, which simultaneously changes the plasma boundary and coil shapes. This is beyond the scope of the current study.

\subsection{US111 with an umbilic coil current in the same direction as the plasma current (co-$I$ case)}
\label{subsec:co-I}
\begin{figure}
    \centering
    \begin{subfigure}[b]{0.325\textwidth}
    \centering
        \includegraphics[width=\textwidth, trim={0mm 20mm 0mm 2mm}, clip]{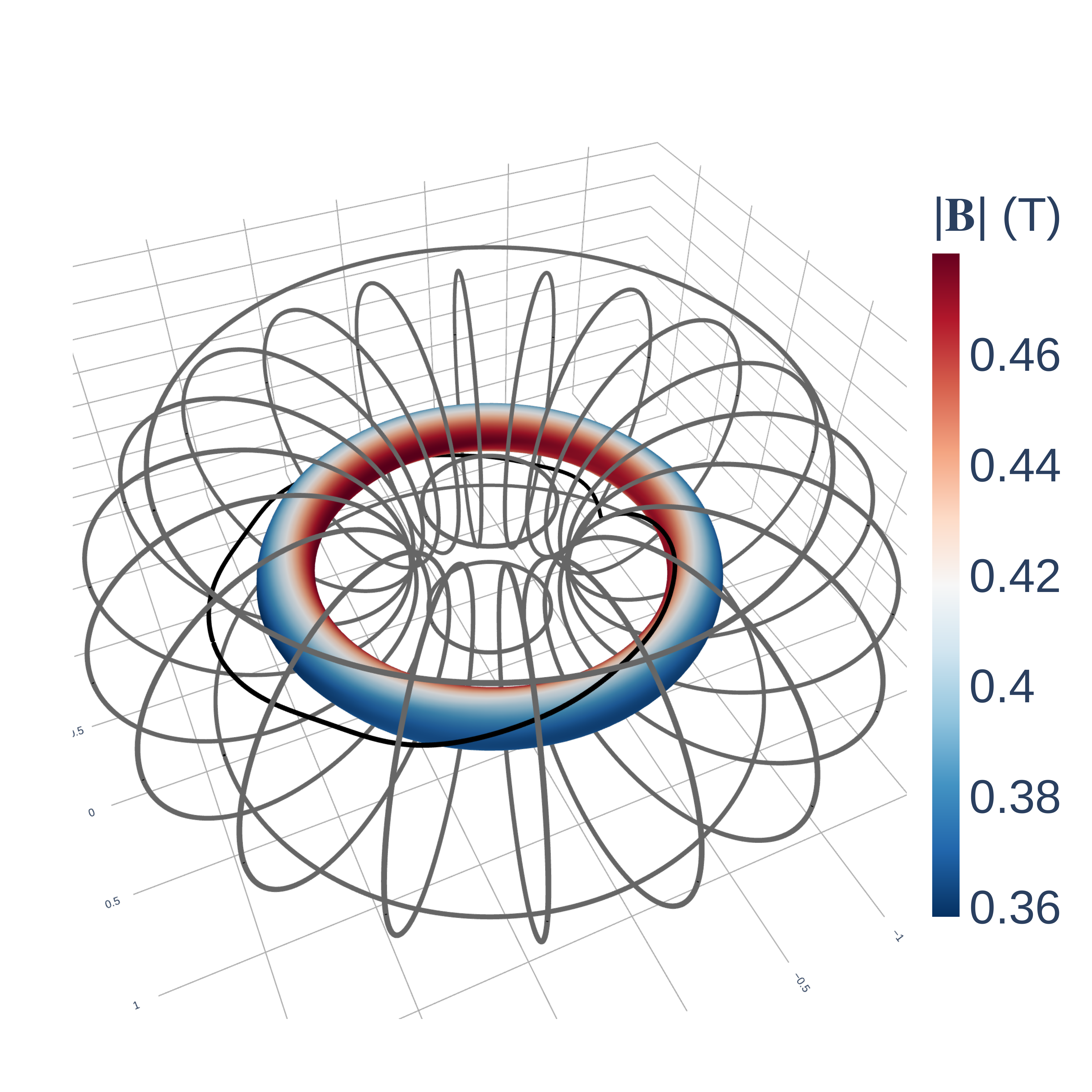}
        \caption{Mod. HBT-EP full coilset}
    \end{subfigure}
    \hspace*{-1mm}
    \begin{subfigure}[b]{0.31\textwidth}
        \captionsetup{margin={-0mm,0cm}}
        \centering
        \includegraphics[width=1.05\textwidth, trim={5mm 20mm 20mm 2mm}, clip]{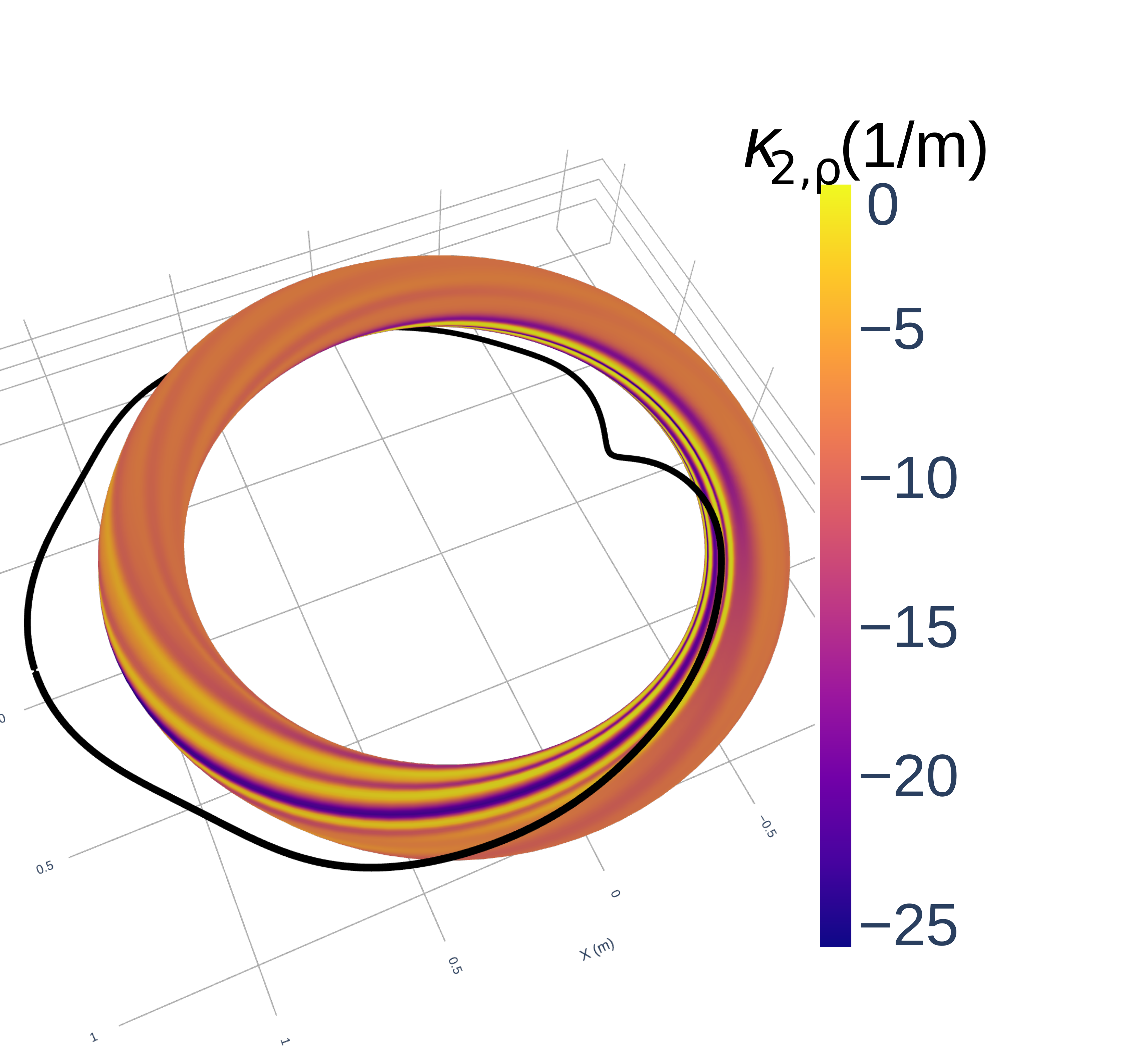}
        \caption{$\kappa_{2, \rho}$ \& umbilic coil}
    \end{subfigure}
    \hspace*{-4mm}
    \begin{subfigure}[b]{0.29\textwidth}
        \captionsetup{margin={0mm,0cm}}
        \centering
        \includegraphics[width=0.95\textwidth, trim={0mm 2mm 30mm -25mm}, clip]{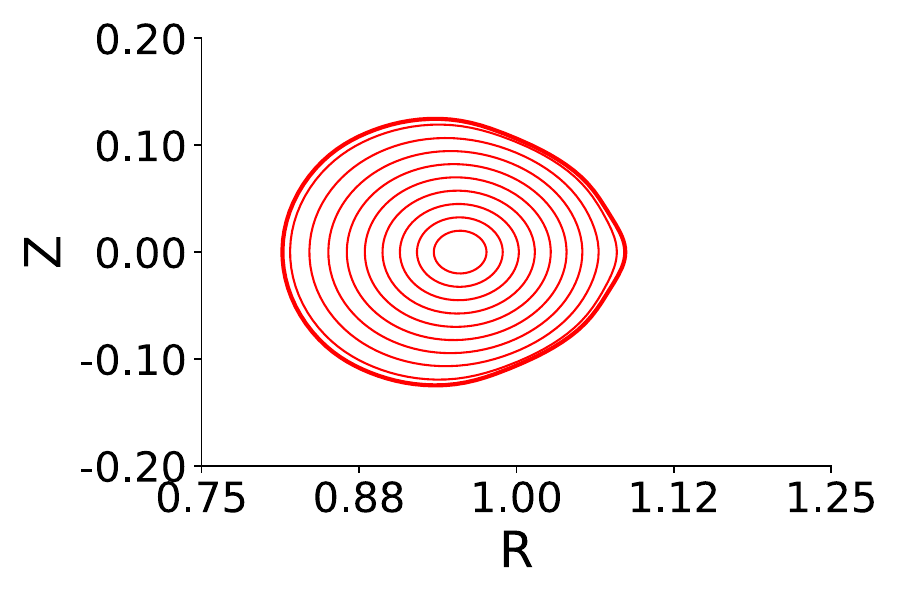}\\[1mm]
        \caption{X-section \& umbilic coil}
    \end{subfigure}
    \caption{Coilset and free-boundary solution for the case where the current in the umbilic coil is in the same direction as the plasma current. The current carried by the umbilic coil $I_{\mathrm{umbilic}} = -3.1kA$. All coils but the umbilic coil are axisymmetric.}
\label{fig:HBT-EP-reverse-result}
\end{figure}

\begin{figure}
    \centering
    \includegraphics[width=0.37\linewidth]{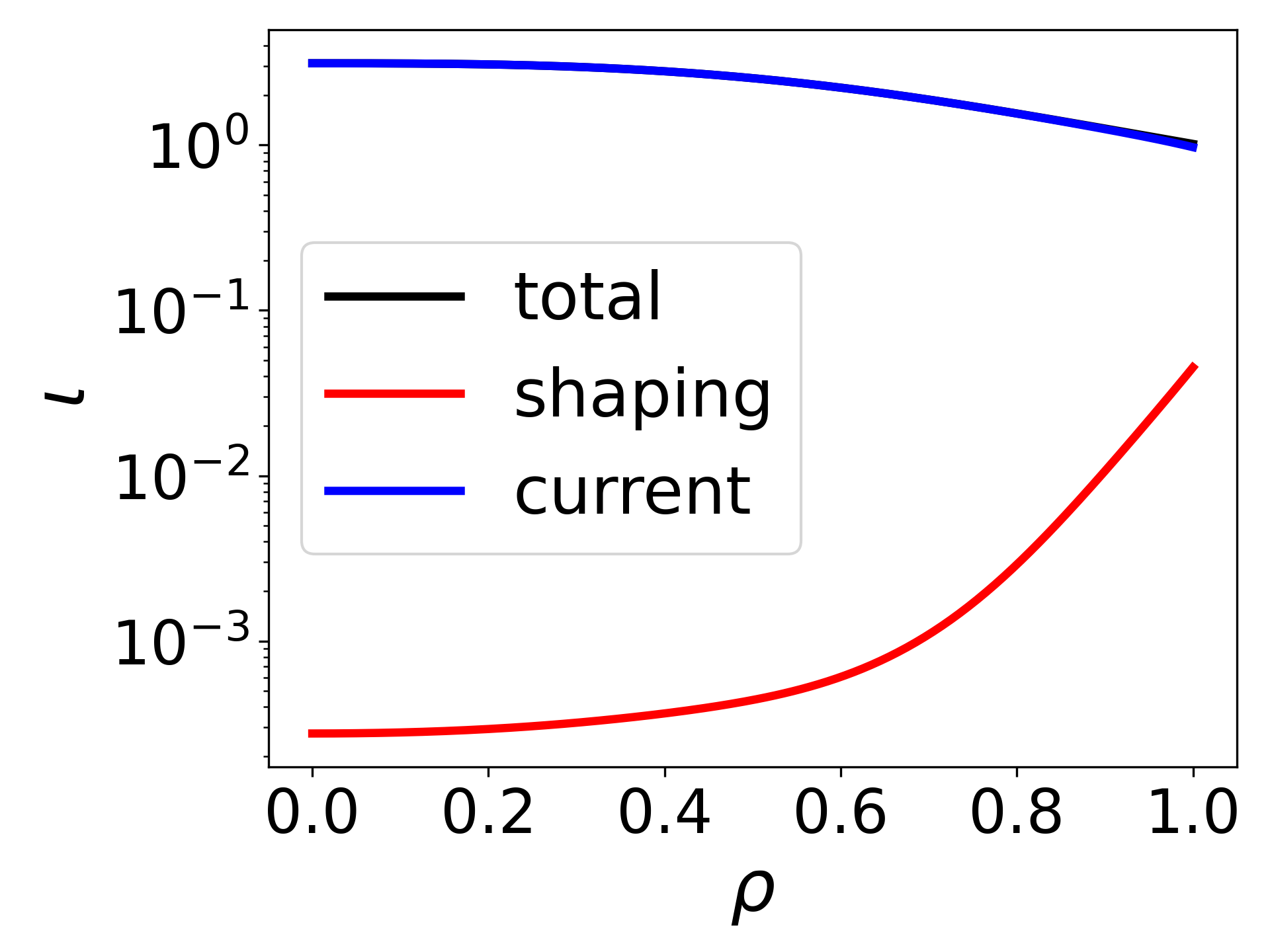}
    \caption{The total rotational transform profiles and its contribution from shaping and plasma current. The non-axisymmetric shaping is purely a consequence of the umbilic coil which contributes to approximately $5\%$ of the total rotational transform in the edge.}
    \label{fig:iota-contribution}
\end{figure}

In a manner similar to the previous section, we create a setup where we modify the HBT-EP experiment and add a $m = n = 1$ ridge to the boundary. However, we do this while ensuring that the toroidal component of the current in the umbilic coil $I_{\mathrm{umbilic}}$ is in the same direction as the plasma current $I = -35.2 kA$. To find the umbilic coil shape, coil current, and the corresponding plasma shape that minimizes the normal field error $\bm{B}\cdot \hat{\bm{n}}$ and the total pressure across the plasma boundary, we combine the objectives defined in~\eqref{eqn:overall-objective} and~\eqref{eqn:second-stage-penalty} in a process known as single-stage optimization. We solve
\begin{equation}
    \min (\mathcal{F}_{\mathrm{stage-one}} + \mathcal{F}_{\mathrm{stage-two}}), \quad \textrm{s.t.} \, \bm{\nabla}\left(\mu_0 p + \frac{B^2}{2}\right) - \bm{B}\cdot\bm{\nabla}\bm{B} =  0,\,  \psi_{\mathrm{b}} = \psi_{\mathrm{b}0}, p = p_{0}(\psi), I = I_0(\psi)
\end{equation}
where $\psi_{\mathrm{b}0}$ is the user-provided enclosed toroidal flux by the boundary, $p_0(\psi)$ and $I_0(\psi)$ are the pressure and toroidal current profiles. The optimizable set of parameters \newline $\hat{\bm{p}} \in \{R_{\mathrm{b}, mn}, Z_{\mathrm{b}, mn}, a_k, X_n, Y_n, Z_n, I_{\mathrm{umbilic}}\}$, where $X_n, Y_n, Z_n$ are the Fourier coefficients that determine the shape of the umbilic coil. The coilset and free-boundary solution are presented in figure~\ref{fig:HBT-EP-reverse-result}.
After optimization, the normalized $\mathrm{max}(\bm{B}\cdot \hat{\bm{n}}) < 0.5\%$ and the total pressure difference is $< 0.1\%$. Because we allow the boundary shape to change, we have less control over the curvature along the umbilic edge, so there is a lot of variation. The curvature along the edge lies in the range $\kappa_{2, \rho} \in [-25, -13]\, m^{-1}$ with the average curvature $\mathrm{avg}(\kappa_{2, \rho}) = -8 m^{-1}$. The neoclassical transport coefficient ${\epsilon_{\mathrm{eff}}}^{3/2} < 0.001$ implies that the magnetic field is still omnigenous.
Finally, we calculate the rotational transform profile of the optimized equilibrium to quantify the effect of the umbilic coil in figure~\ref{fig:iota-contribution}.

The effect on the rotational transform profile due to the umbilic coil rises sharply as we move towards the edge. Therefore it is possible that outside the plasma boundary the umbilic coil is able to affect the total magnetic field in such a way that it forms an X-point.

\section{Conclusions and future work}
\label{sec:fin}
This paper explored the physical properties of umbilic stellarators. In section~\ref{sec:UToL}, we defined the umbilic parametrization, approximated it using~\texttt{DESC} and solved the ideal MHD equation inside the enclosed volume. In section~\ref{sec:eq-only-optimization}, we developed vacuum and finite-$\beta$ stellarators by simultaneously optimizing the umbilic edge and the boundary shape using the~\texttt{DESC} optimization suite. We designed coils for the vacuum case, and traced fieldlines in the edge of the US131 stellarator before and after adding a current source at the magnetic axis, to simulate a plasma fluctuation. Even though we do not see a perfectly sharp X-point (X-line) and well-defined fieldline structure like a tokamak, the high-curvature edges could be a favorable location to place divertors and are robust to internal plasma fluctuations. We found that the optimized omnigenous umbilic designs do not possess a specific helicity regardless of the low-ripple transport, which could be a way to generate piecewise omnigenous equilibria. As a final demonstration, in section~\ref{sec:HBT-coil}, we propose an experiment to modify the Columbia HBT-EP experiment to an umbilic design by adding a $m=1, n=1$ helical coil. Using the~\texttt{DESC} free boundary solver, we demonstrate that the umbilic coils are able to cause the originally circular plasma to bulge and form a high-curvature ridge while maintaining omnigenity. By changing the direction of the current in the umbilic coil, we may be able to get an X-point, thereby converting the limited HBT-EP tokamak into a diverted stellarator.

The technique of simultaneously optimizing a curve and a surface to control the magnitude and position of a sharp edge along a surface extends beyond umbilic stellarators. One can use this method to place and design ridges or regions of high curvature which may be useful for non-resonant divertors~\citep{bader2018minimum}, as well as these ridges creating preferential pathways for the field lines to exit the plasma~\citep{bader2017hsx}. This can be achieved by multiplying a custom function $g(\theta)$ to the objective $f_{\mathrm{umbilic}}$ such that $g(\theta) = 1\quad  \forall \quad \theta \in \bigcup_{i} [\theta_{i0}, \theta_{i1}]$ and $g(\theta) = 0$ elsewhere along the umbilic curve, with $\theta_{i0}, \theta_{i1}$ being optimizable parameters describing the lower and upper limits of regions where low curvature is penalized. This will create a set of continuous but disconnected ridges, just as we observe in non-resonant divertor concepts. This functionality is available in~\texttt{DESC}. For an island divertor concept, the rotational transform on the boundary can be fine-tuned using umbilic coils, which could be used to reduce the sensitivity of X-points in the island divertor structure due to changes in the plasma boundary rotational transform. This idea is similar to that of divertor coils in tokamaks and would lead to a predictable strike point pattern. 

Future work would involve using multiple umbilic coils to ensure plasma stability against low-$n$ modes ($n = 0$ family) and increase the curvature of the ridge, similar to the work on tokamak separatrix by~\citet{webster2009techniques}, understanding the strike point pattern and field line structure outside the plasma boundary or designing umbilic coils for a case similar to the finite-$\beta$ $\rm{US}252$ stellarator in section~\ref{subsec:UT135}. It is also important to include MHD and kinetic stability metrics in the optimization process, especially near the boundary as has been done for tokamaks by~\citet{zheng2025x,bishop1986stability} as stability properties can be sensitive close to the high-curvature plasma boundary.

\vspace*{5mm}
\textbf{Acknowledgements}
One of the authors, R. G. enjoyed discussions with M. Zarnstroff, J. Levesque, R. Davies, J. L. Velasco, B. Jang, and gratefully acknowledges the help from S. Lazerson installing the~\texttt{FIELDLINES} code. We also express our gratitude to the referees for their constructive feedback that has improved the overall quality of this paper.

\textbf{Funding}
This work is funded through the SciDAC program by the US Department of Energy,
Office of Fusion Energy Science, and Office of Advanced Scientific Computing Research
under contract No. DE-AC02-09CH11466, the Hidden Symmetries grant from the Simons Foundation/SFARI (560651), 
and Microsoft Azure Cloud Computing mini-grant awarded by the Center for Statistics and Machine Learning (CSML), Princeton University. This research also used the Della cluster at Princeton University. 

\textbf{Data availability statement} All the data, analysis, and post-processing files are freely available in~\citet{gaur_2025_15355215}.

\textbf{Declaration of interests} Matt Landreman is a consultant for Type One Energy

\appendix
\section{Calculating principal curvatures}
\label{sec:appendix-curvature}
The second principal curvature at a given point on a surface is defined as a 
\begin{equation}
\centering
\kappa_{2, \rho} = \min\left\{k:\det\left[\begin{array}{cc}
    L_{\mathrm{sff}, \rho} - k E & M_{\mathrm{sff}, \rho} - k F \\
    M_{\mathrm{sff}, \rho} - k F & N_{\mathrm{sff}, \rho} - k G 
\end{array}\right] = 0\right\}
\end{equation}
where
\begin{equation}
\begin{split}
    E &= \frac{\partial \bm{x}}{\partial \zeta} \cdot \frac{\partial \bm{x}}{\partial \zeta},\quad F = \frac{\partial \bm{x}}{\partial \zeta} \cdot \frac{\partial \bm{x}}{\partial \theta},\quad  G = \frac{\partial \bm{x}}{\partial \theta} \cdot \frac{\partial \bm{x}}{\partial \theta}, \\
    L_{\mathrm{sff}, \rho} &= \frac{\partial^2 \bm{x}}{\partial \theta^2} \cdot \frac{\bm{\nabla}\rho}{|\bm{\nabla}\rho|}, \quad   M_{\mathrm{sff}, \rho} = \frac{\partial^2 \bm{x}}{\partial \theta \partial \zeta} \cdot \frac{\bm{\nabla}\rho}{|\bm{\nabla}\rho|}, \quad N_{\mathrm{sff}, \rho} = \frac{\partial^2 \bm{x}}{\partial \zeta^2} \cdot \frac{\bm{\nabla}\rho}{|\bm{\nabla}\rho|}
\end{split}
\end{equation}
are the metric coefficients corresponding to the first and second fundamental forms. The user-specified values $\kappa_{2, \mathrm{bound1}}, \kappa_{2, \mathrm{bound2}}$ define the limits on the curvature. Using this objective, we impose a maximum and minimum value on $\kappa_{2, \rho}$ that helps us impose a sharp curvature along the umbilic edge.

\section{Form of various objectives}
\label{sec:objectives-FoMs-defn}
All the objectives with bounds are defined as
\begin{eqnarray}
    f_{\mathrm{obj}} = \mathrm{ReLU}(\max(\rm{obj} - \rm{obj}_{\mathrm{bound1}}, \rm{obj} - \rm{obj}_{\mathrm{bound2}})).
\end{eqnarray}
In this paper, such a term is used to penalize the aspect ratio, rotational transform, and the second principal curvature $\kappa_{2, \rho}$ along the umbilic edge. 

For the effective ripple objective explained in Unalmis~\textit{et al.}~\citep{unalmis2024spectrally} and the omnigenity objective implemented by Dudt~\textit{et al.}~\citep{dudt2024magnetic}, we do not use any bounds and try to reduce the magnitude of these quantities as much as possible. The exact definitions are given in the respective papers.

\section{Piecewise omnigenity in umbilic configurations}
\label{sec:pwO_plots}
\begin{figure}
    \centering
    \begin{subfigure}[b]{0.34\textwidth}
    \centering
        \includegraphics[width=\textwidth, trim={0mm 0mm 0mm 2mm}, clip]{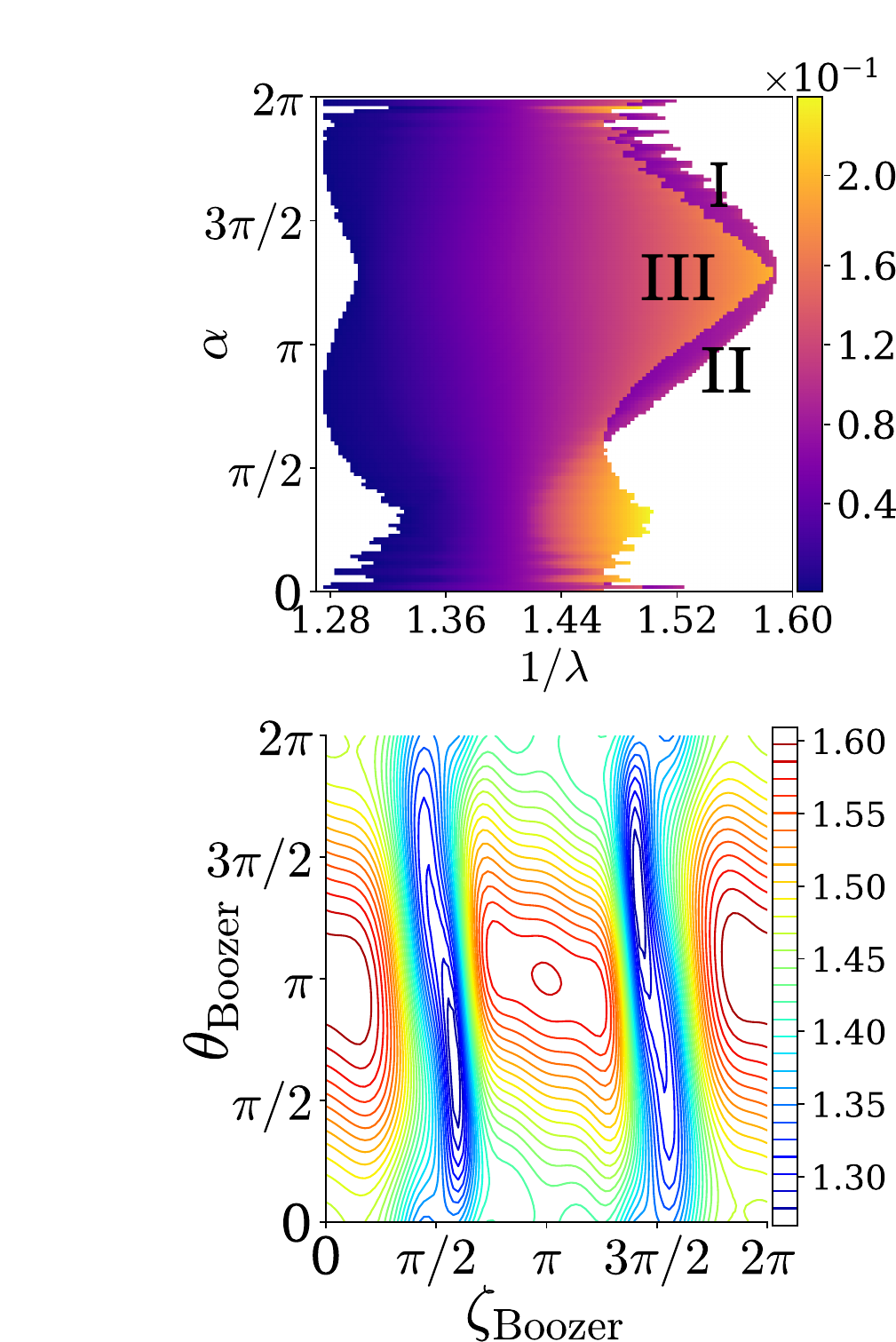}
        \caption{US131 $(\rho = 0.5)$}
    \end{subfigure}
    \begin{subfigure}[b]{0.304\textwidth}
        \centering
        \includegraphics[width=1.0\textwidth, trim={2mm 0mm 0mm 0mm}, clip]{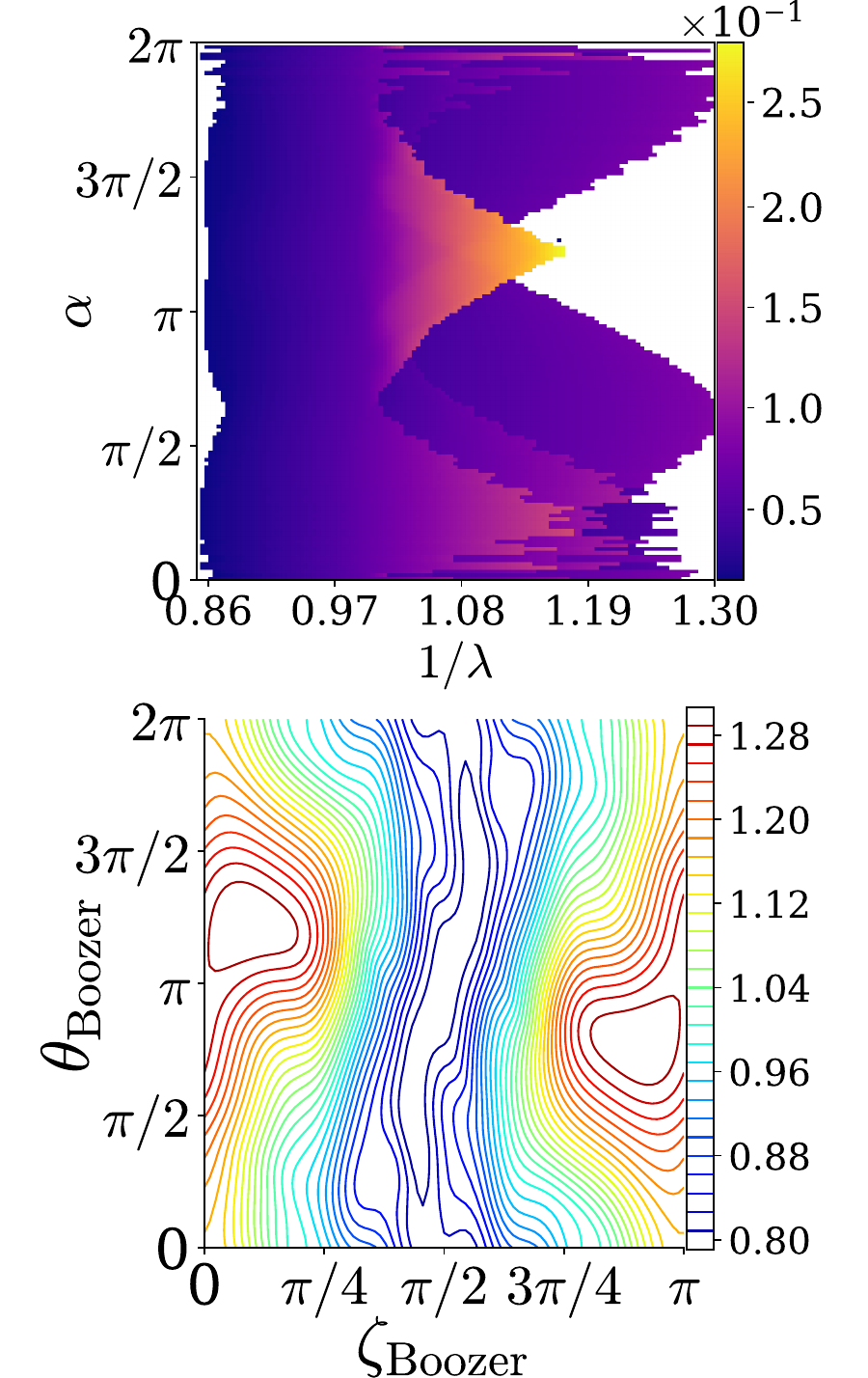}
        \caption{US252 $(\rho = 0.75)$}
    \end{subfigure}
    \begin{subfigure}[b]{0.2755\textwidth}
        \captionsetup{margin={-4mm,0cm}}
        \centering
        \includegraphics[width=1.0\textwidth, trim={0mm -1mm 0mm 0mm}, clip]{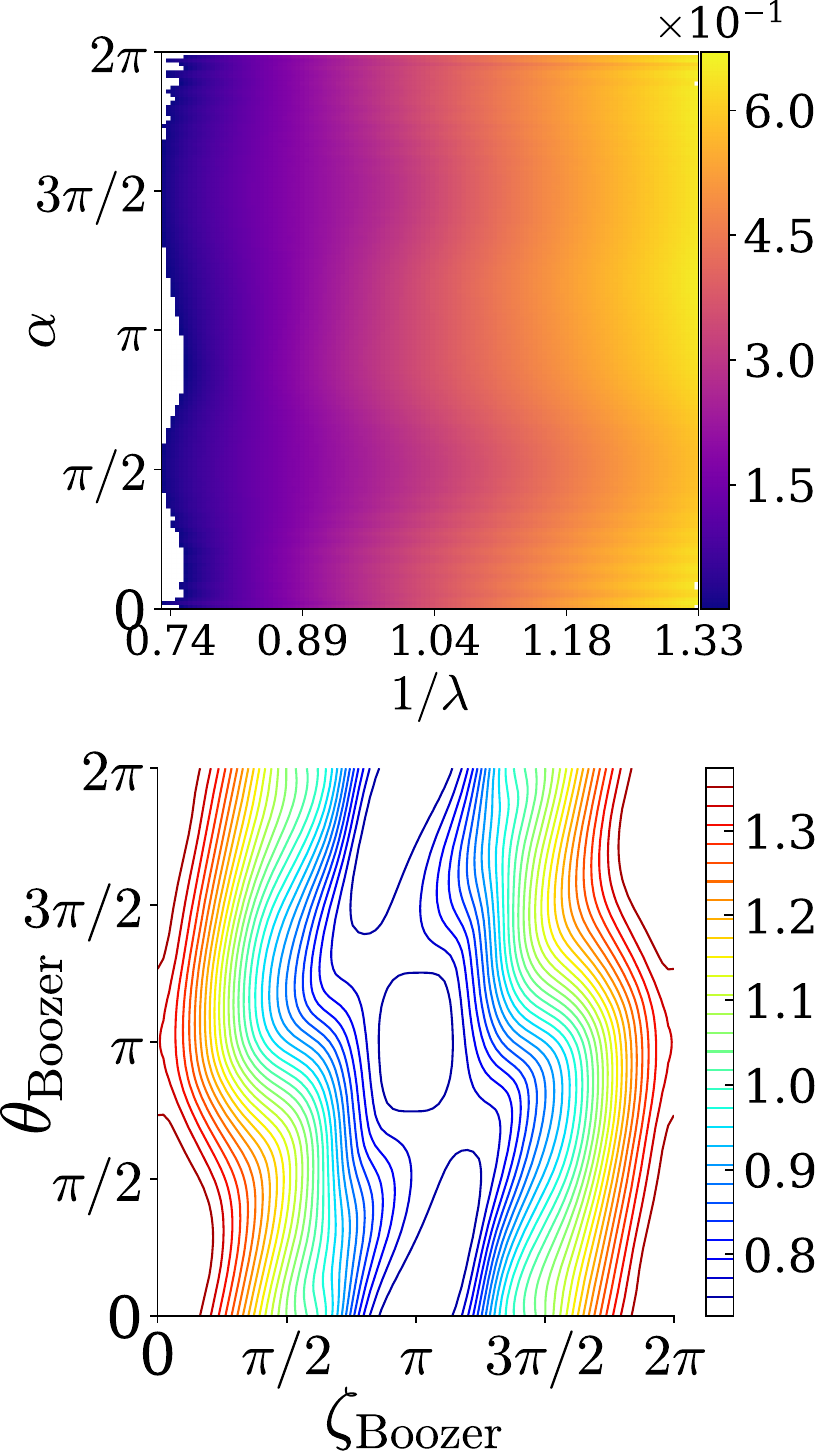}\\[-0.0mm]
        \caption{$\NFP\!=\!1,\! A\!=\!5$ OP$(\rho\! =\!1)$}
    \end{subfigure}
    \caption{Plots of the second adiabatic invariant $\mathcal{J}_{\parallel}(\alpha, 1/\lambda)$ and $|B|$ contours for three equilibria with $\epsilon_{\mathrm{eff}}^{3/2} \approx 0.001$. For a close-to-perfect omnigenous equilibrium $\partial_{\alpha}\mathcal{J}_{\parallel}$ close to zero for most values of $1/\lambda$, as seen in~\textit{(c)}. However, for~\textit{(a)} and~\textit{(b)} there are multiple distinct values of $\mathcal{J}_{\parallel}$, discontinuous in $\alpha$, a necessary conditions for piecewise omnigenity.}
\label{fig:pwO-plots}
\end{figure}

In this appendix, we will plot the normalized second adiabatic invariant $\mathcal{J}_{\parallel} = \int_{l_1}^{l_2} dl\, v_{\parallel}/(2\pi R_0)$ as a function of the field line label $\alpha$ and the inverse pitch angle $1/\lambda \equiv E/\mu$. We calculate $\mathcal{J}_{\parallel}(\alpha, 1/\lambda)$ on a single flux surface on which the neoclassical transport coefficient $\epsilon_{\mathrm{eff}}^{3/2} \leq 0.001$. This is done for both the US131 vacuum and US252 finite-beta equilibria presented in sections~\ref{subsec:vacuum-UToL} and~\ref{subsec:UT135}, respectively. To highlight their difference, we also plot $\mathcal{J}_{\parallel}$ for a poloidally omnigenous (or QI) $\NFP=1$ vacuum equilibrium taken from the~\texttt{DESC} omnigenity database~\citep{Gaur-DPP23}. Comparison is shown in figure~\ref{fig:pwO-plots}.

For a perfect, non-piecewise omnigenous stellarator $\partial_{\alpha}(\mathcal{J}_{\parallel}) = 0$, implying that $\mathcal{J}_{\parallel}$ is continuous and constant in $\alpha$ for almost every $1/\lambda$. Moreover, the distinct $\mathcal{J}_{\parallel}$ can be written as a linear sum such that the total $\mathcal{J}_{\parallel}$ for a value of the inverse pitch vanishes. However, for the umbilic configurations in this paper, especially the US131 vacuum stellarator, despite a low effective ripple (good omnigenity), we clearly see that $\mathcal{J}_{\parallel}$ is discontinuous in $\alpha$ for most of the inverse pitch values and $\tilde{\mathcal{J}}_{\parallel, \mathrm{I}} \approx \tilde{\mathcal{J}}_{\parallel, \mathrm{II}} \approx 0.5 \tilde{\mathcal{J}}_{\parallel, \mathrm{III}}$. This supports our claim that the umbilic stellarator US131 is piecewise omnigenous, especially at large values of the inverse pitch. US252 seems to be piecewise omnigenous but the structure of $\tilde{\mathcal{J}_{\parallel}}$ is more complicated than the one discussed by~\citet{velasco2024piecewise}.

\section{Details of the HBT-EP equilibrium}
\label{app:HBT-EP}
\begin{figure}
    \centering
    \begin{subfigure}[b]{0.34\textwidth}
    \centering
        \includegraphics[width=\textwidth, trim={0mm 0mm 1mm 0mm}, clip]{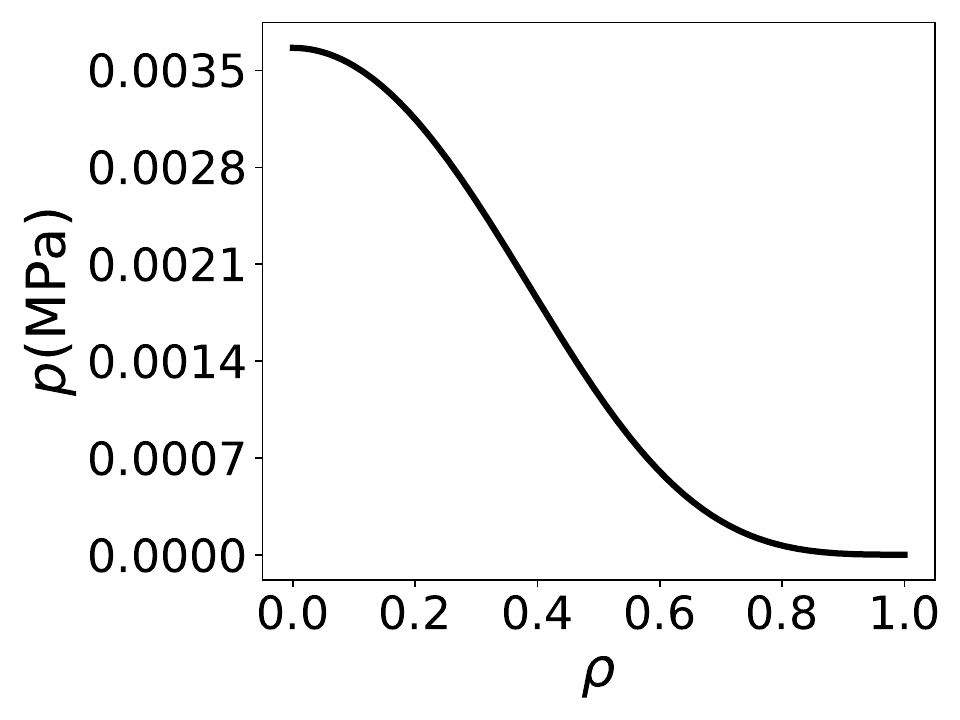}
        \caption{Pressure}
    \end{subfigure}
    \begin{subfigure}[b]{0.34\textwidth}
        \centering
        \includegraphics[width=1.0\textwidth, trim={0mm 0mm 1mm 0mm}, clip]{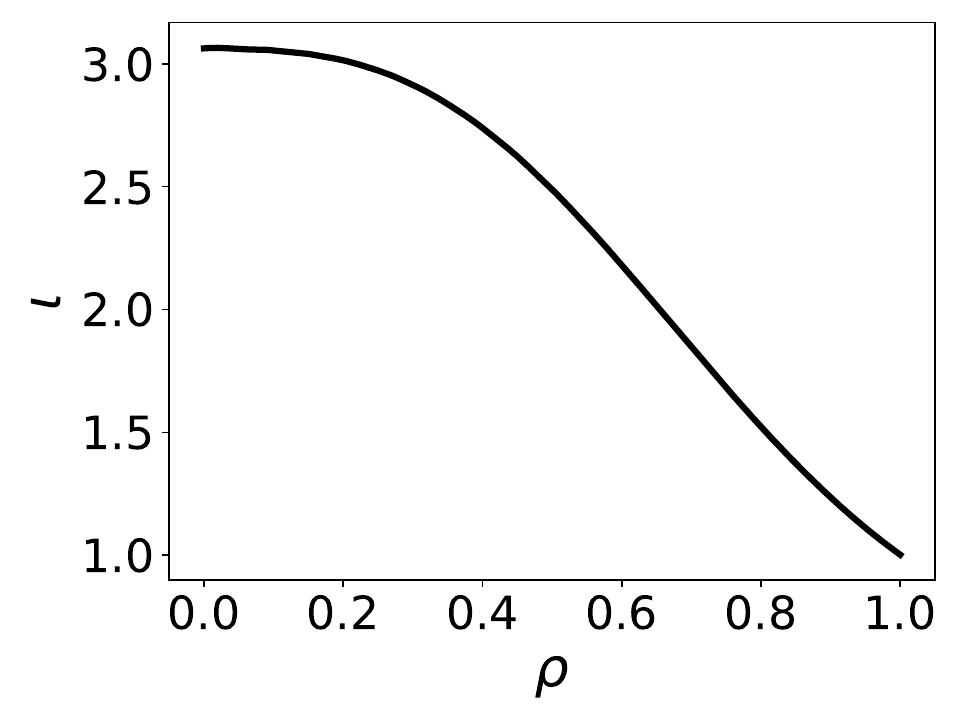}
        \caption{Rotational Transform}
    \end{subfigure}
    \caption{Profiles of the initial HBT-EP equilibrium.}
\label{fig:HBT-EP-profiles}
\end{figure}

The original HBT-EP equilibrium is a shifted-circle equilibrium with a major radius $R_0 = 0.94 m$, minor radius $a_0 = 0.125 m$ (aspect ratio $A = 7.52$) with a $\beta = 0.5\%$ and a plasma current $I 
 = -35.2 kA$. The pressure and rotational transform profiles are given in figure~\ref{fig:HBT-EP-profiles}. The initial equilibrium was tested to be stable against $n = 1$ kink modes using the~\texttt{DCON-2D}~\citep{glasser2016direct} stability code. 

\bibliographystyle{jpp}
\bibliography{jpp-instructions}

\end{document}